\documentclass[fleqn, usenatbib]{mnras}

\usepackage{newtxtext}

\usepackage[T1]{fontenc}
\usepackage{ae,aecompl}

\usepackage{graphicx}	
\usepackage{amsmath}	
\usepackage{amssymb}	
\usepackage{subfig}
\usepackage[font=footnotesize]{caption}
\usepackage{wrapfig}
\usepackage{gensymb}
\usepackage{multirow}
\usepackage{tabularx}
\usepackage{csquotes}
\usepackage{float}
\usepackage{lscape}
\usepackage{stackengine}
\usepackage[dvipsnames]{xcolor}
\usepackage{tablefootnote}
\usepackage{enumitem}
\usepackage{longtable}
\usepackage{threeparttablex}
\usepackage{hhline}	
\usepackage{pdflscape}	
\usepackage{comment}
\usepackage{breqn}

\def\maxififth{MAXI~J1535--571}
\def\maxithirt{MAXI~J1348--630}
\def\maxieight{MAXI~J1820$+$070}
\def\xte{XTE~J1550--564}
\def\xtej{XTE~J1752--223}
\def\hh{H1743--322}
\def\gx{GX~339--4}
\def\grs{GRS~1915$+$105}

\title[Large-scale jets from BH XRBs]{Constraining the physical properties of large-scale jets from black hole X-ray binaries and their impact on the local environment with blast-wave dynamical models}

\author[F. Carotenuto et al.]{F. Carotenuto,$^{1}$\thanks{E-mail: francesco.carotenuto@physics.ox.ac.uk}
R. Fender,$^{1}$
A. J. Tetarenko,$^{2,3\thanks{Former  NASA Einstein Fellow}}$
S. Corbel,$^{4,5}$
A. A. Zdziarski,$^{6}$
G. Shaik,$^{7}$
A. J. Cooper,$^{1}$
\newauthor
I. Di Palma$^{7,8}$
\\
$^{1}$Astrophysics, Department of Physics, University of Oxford, Keble Road, Oxford OX1 3RH, UK\\
$^{2}$Department of Physics and Astronomy, University of Lethbridge, Lethbridge, Alberta, T1K 3M4, Canada\\
$^{3}$Department of Physics \& Astronomy, Texas Tech University, Lubbock, TX 79409-1051, USA\\
$^{4}$AIM, CEA, CNRS, Universit\'{e} Paris Cité, Universit\'{e} Paris-Saclay, F-91191 Gif-sur-Yvette, France\\
$^{5}$Observatoire Radioastronomique de Nan\c cay, Observatoire de Paris, PSL Research University, CNRS, Univ. Orl\'eans, 18330 Nan\c cay, France\\
$^{6}$Nicolaus Copernicus Astronomical Center, Polish Academy of Sciences, Bartycka 18, PL-00-716 Warszawa, Poland\\
$^{7}$Università di Roma “La Sapienza”, I-00185 Roma, Italy\\
$^{8}$Istituto Nazionale di Fisica Nucleare, Sezione di Roma, P.\ le Aldo Moro 2, I-00185 Rome, Italy\\
}
\date{Accepted XXX. Received YYY; in original form ZZZ}

\pubyear{2024}

\begin{document}
\label{firstpage}
\pagerange{\pageref{firstpage}--\pageref{lastpage}}
\maketitle

\begin{abstract}
\noindent
Relativistic discrete ejecta launched by black hole X-ray binaries (BH XRBs) can be observed to propagate up to parsec-scales from the central object. Observing the final deceleration phase of these jets is crucial to estimate their physical parameters and to reconstruct their full trajectory, with implications for the jet powering mechanism, composition and formation. In this paper we present the results of the modelling of the motion of the ejecta from three BH XRBs: \maxieight{}, \maxififth{} and \xtej{}, for which high-resolution radio and X-ray observations of jets propagating up to $\sim$15 arcsec ($\sim$0.6 pc at 3 kpc) from the core have been published in the recent years. For each jet, we modeled its entire motion  with a dynamical blast-wave model, inferring robust values for the jet Lorentz factor, inclination angle and ejection time. Under several assumptions associated to the ejection duration, the jet opening angle and the available accretion power, we are able to derive stringent constraints on the maximum jet kinetic energy for each source (between $10^{43}$ and $10^{44}$ erg, including also \hh{}), as well as placing interesting upper limits on the density of the ISM through which the jets are propagating (from $n_{\rm ISM} \lesssim 0.4$ cm$^{-3}$ down to $n_{\rm ISM} \lesssim 10^{-4}$ cm$^{-3}$). Overall, our results highlight the potential of applying models derived from gamma-ray bursts to the physics of jets from BH XRBs and support the emerging picture of these sources as preferentially embedded in low-density environments.
\end{abstract}



\section{Introduction}
\label{sec:Introduction}

Relativistic jets appear as an ubiquitous feature among accreting black holes (BH) in the Universe, from supermassive BHs in active galactic nuclei (AGN) to stellar-mass BHs in galactic X-ray binaries (XRBs). Highly relativistic jets are also produced in energetic transients events, such as gamma-ray bursts (GRBs) and tidal disruption events (TDEs), a large fraction of which are believed to be powered by accreting BHs. The short timescales of evolution (days to weeks) and the relative proximity of BH XRBs make them ideal targets on which to study the properties of relativistic jets \citep{Fender2006, Romero_2017}, some of which appear to also be scale-invariant, and thus valid for all accreting BHs \citep{Kording_2005}. In BH XRBs, different jets are produced in different phases of the outburst \citep{Corbel_2004, Fender_belloni_gallo}. Compact jets, causally connected to the accretion flow, are observed during the hard spectral state (see \citealt{Remillard_xrb} and \citealt{Homan_belloni} for a review on spectral states), as they emit self-absorbed synchrotron radiation, which dominates in the radio through near-infrared \citep{Corbel_2000, Fender_2001, Markoff_2001, Corbel2002, Russell_2013b}. On the other hand, discrete ejecta are observed to be launched during transitions between the hard and the soft states, producing strong multi-wavelength flares during which the synchrotron emission is initially self-absorbed and then optically thin at radio wavelengths (e.g.\ \citealt{Tetarenko2017}). These components consist of bipolar blobs of plasma that travel away from the core, often at apparently superluminal speeds, and might be considered as less-relativistic analogs of what is observed in AGN (e.g.\ \citealt{Marscher_2002, Gomez_2008}). As of today, spatially resolved discrete ejecta have been observed with radio-interferometric observations in 15 sources: GRS~1915$+$105 \citep{Mirabel1994}, GRO~J1655--40 \citep{Hjellming_1995, Tingay_1995}, Cyg~X-3 \citep{Mioduszewski}, GX~339-4 \citep{Gallo_2004}, \xte{} \citep{Hannikainen_2001, Corbel2002_xte}, \xtej{} \citep{Yang2010, MJ_2011}, \hh{} \citep{Corbel2005_h17, Miller-Jones_h1743}, XTE~J1859$+$226 \citep{Rushton}, \maxififth{} \citep{Russell_1535}, V404~Cyg \citep{Miller-Jones2019}, \maxieight{} \citep{Bright, Espinasse_xray}, \maxithirt{} \citep{Carotenuto2021}, EXO~J1846--031 \citep{Williams_2022}, MAXI~J1803--298 \citep{Wood_2023} and MAXI~J1848--105 \citep{Bahramian_2023}. This sample represents around $20\%$ of the current population of confirmed and candidate BH XRBs, which are, however, all believed to produce jets \citep{Watchdog, Corral_santana}. In case of non-detection, this is likely due to source being too far, or with an unfavorable inclination angle (due to the effect of Doppler boosting) or to a failed transition from the hard to the soft state, which was found to happen, approximately, for a third of the observed outbursts \citep{Alabarta_2021}. For sources that do display hard-to-soft state transitions, optically thin radio flares can be always detected with an adequate radio monitoring. In some cases, discrete ejecta can propagate up to parsec scales far from the core, displaying re-brightenings and deceleration phases likely due to the interaction with the interstellar medium (ISM), which also result in the production of broadband synchrotron radiation (radio to X-rays) from \textit{in-situ} particle acceleration, up to TeV energies \citep{Corbel2002_xte, Corbel2005_h17, Migliori2017, Espinasse_xray, Carotenuto2021}.

Despite the wealth of multi-wavelength observations collected over the recent years, multiple aspects related to the formation, evolution and overall physics of these jets remain unclear. For instance, the powering mechanism of the jets is still an open problem, as jets could be powered from the extraction of energy from a spinning BH \citep{Blandford_1977} or from its accretion disk \citep{Blandford_1982}, or from a combination of the two, as suggested by general relativistic magnetohydrodynamic (GRMHD) simulations \citep{McKinney_2006} and recent Event Horizon Telescope observations probing a possible light spine vs.\ massive sheath jet structure \citep{Janssen_2021}. The plasma composition, either baryonic or purely leptonic, is also unknown, and it is notoriously difficult to constrain as most jets display only a simple featureless synchrotron spectrum \citep{Fender2006}. Moreover, while radio/infrared timing techniques are opening a new window on the physical parameters of compact jets \citep{Casella_2010, Tetarenko_2019, Tetarenko_2021, Zdziarski_2022}, and recent results found evidence for a luminosity dependence of their properties \citep{Prabu_2023}, we still lack precise constraints on the physical parameters of discrete ejecta, such as their mass, speed, energy and volume. In particular, a key open problem is the quantification of their total energy content. Measuring the jet's energy is of prime importance not only to estimate the balance between inflows and outflows in BH XRBs, but also because of the implications on the jet composition, powering mechanism and impact on the surrounding environment (e.g.\ \citealt{Fender_balance}). 

The total energy (internal plus kinetic) of discrete ejecta can be estimated with different approaches. First, given the jet synchrotron emission, it is possible to infer the internal energy of the plasma that is required to produce the observed radiation by relying on the knowledge of the size of the emitting region and of the source distance, while assuming equipartition conditions \citep{Longair}. The size of the emitting region can be most easily estimated by directly resolving the plasmon with radio or X-ray observations (e.g.\ \citealt{Rushton}). When this is not possible, the synchrotron-emitting region size can be estimated through the detection of the radio spectral peak due to synchrotron-self absorption (e.g.\ \citealt{Fender_2019_equipartition} for BH XRBs), or it can be computed assuming a jet expansion speed and an ejection timescale (usually the duration of the rise of the radio flare at the jet’s launch), although this may largely underestimate the jet's internal energy \citep{Bright, Carotenuto_2022}. An additional way of measuring the jet size relies on simultaneous radio-interferometric observations of the ejecta at the same frequency, but with different angular resolutions probing different spatial scales. By measuring the percentage of flux resolved out between the observations, it is possible to infer the size of the emitting region and subsequently the jet internal energy \citep{Bright}. Alternatively, it is possible to identify jet-produced structures in ISM and then use them as calorimeters to measure the mechanical power, and consequently the kinetic energy that the jets need to deposit in those structures in order to create and sustain them \citep{Gallo_2005, Russell_2007, Tetarenko_alma, Tetarenko_2020_ism}. 

Independently, focusing on the kinematics of these jets and covering their full trajectory allows us to obtain a complete dataset  (angular separation vs.\ time) of their evolution, which can later be used to test physical models for the jet propagation in the ISM. The application of these models, which are mostly derived from the physics of GRBs \citep{Wang_model}, can yield important constraints on multiple physical parameters of the ejecta, such as their Lorentz factor, mass, inclination angle, ejection time and kinetic energy \citep{Wang_model, Hao, Steiner_xte}. In particular, covering the jet deceleration phase with dense monitoring campaigns can significantly improve the constraints from these models \citep{Carotenuto_2022}. Furthermore, modelling the jet motion is also of prime importance to precisely constrain the time of ejection, which is fundamental to put the jet launch in context with other multi-wavelength observational signatures, such as radio flares, X-ray spectral changes and X-ray quasi-periodic oscillations (QPOs, e.g.\ \citealt{Ingram_2019}). This allows us to ultimately obtain a comprehensive view of the source evolution during the state transition, with a special focus on the hot corona of electrons that surrounds the BH, which is responsible for the hard X-ray emission and it is thought to be intimately connected to the jet (e.g.\ \citealt{Rodriguez_2003, Markoff_corona, Kara_2019, Mendez_2022, ixpe_ingram}). 

Tracing the jet motion has also turned out to be to especially useful to make use of the jets as probes of the environment surrounding BH XRBs. In fact, different works considering the propagation of ejecta in the ISM have provided strong evidence that BH XRBs are generally located in environments that appear 2-4 orders of magnitude less dense than the canonical Galactic ISM density of 1 particle per cm$^{-3}$ (at least in the direction of the jet propagation), unless these jets are very narrowly collimated, with opening angles $\ll 1\degree$, or extremely energetic, with kinetic energies above $10^{46}$ erg \citep{Heinz_2002, Hao, Carotenuto_2022, Zdziarski_2023}.

The wealth of information that can be extracted from this type of modelling was first shown with the application on the large-scale decelerating jets from the BH XRBs \xte{} \citep{Hao, Steiner_xte}, \hh{} \citep{Steiner_h17}, and, more recently, for \maxithirt{} \citep{Carotenuto_2022, Zdziarski_2023}. In this paper, as a continuation of the work started in \cite{Carotenuto_2022}, we expand the sample of sources that displayed unambiguously decelerating discrete ejecta, and for which such modelling has been applied, to include the jets from the BH XRBs \maxieight{}, \maxififth{} and \xtej{}. These sources displayed resolved, large-scale decelerating jets observed between 2010 and 2018 that were observed to propagate up to $\sim$15 arcsec far from the core \citep{Yang2010, MJ_2011, Yang_2011, Russell_1535, Bright, Espinasse_xray}. However, the jet motion in these sources has only been described with basic phenomenological models, mostly applied in order to constrain the ejection date in relation with the simultaneous X-ray activity of the core. Since the quality of the jet angular separation data justifies the application of a physical model to describe the entire jet evolution, we performed such modelling and we present the detailed results in this paper. 
In particular, we present and discuss new constraints on the jet Lorentz factor, inclination angle, ejection time, as well as upper limits on the maximum energy available to the jets, on the density of the ISM that surrounds the systems and on the mass of the ejecta. For the last part, we also consider the ejecta launched in 2003 by \hh{}, where similar modelling has been already published by \cite{Steiner_h17}.

In Section \ref{sec:Data} we present the sources and the observational data considered for the modelling work, while in Section \ref{sec:The dynamical model} we discuss in detail the dynamical model that we adopted. Then, in Section \ref{sec:Results} we present the results of the application of such model to our data and we discuss our findings in relation to the current understanding of jets from XRBs in Section \ref{sec:Discussion}. Finally, we summarize our conclusions in Section \ref{sec:Conclusions}.

\section{Data}
\label{sec:Data}

The data on the ejecta launched by the three BH XRBs considered in this paper have been already published and are therefore available in the literature.
In the following sections, we present the sources and discuss the data used for this work.

\subsection{\maxieight{}}
\label{sec:1820}

The BH XRB \maxieight{} was discovered by the Monitor of All-sky X-ray Image on board the International Space Station \citep{Matsuoka_maxi} in March 2018 \citep{Kawamuro_2018}, and it was subsequently identified with the optical transient ASASSN-18ey \citep{Denisenko_2018}. \maxieight{} is one of the most well-observed and well-studied BH XRBs in recent years. It harbors a $6.75^{+0.64}_{-0.46} \ M_{\odot}$ BH accreting from a $0.5 \pm 0.1 \ M_{\odot}$ companion star \citep{Torres_2019, Torres_2020, Mikolajewska_2022}. Due to its impressive brightness, primarily in the X-rays \citep{Fabian_2020}, it has been the subject of numerous multi-wavelength observing campaigns across the entire electromagnetic spectrum (e.g.\ \citealt{Shidatsu_2018, Hoang_2019, Bright, Tetarenko_2021, Abe_2022, Cangemi_2023, Echiburu_2024}) yielding a
dataset of extremely high quality for a BH XRB in outburst. A model-independent measurement of the distance of $2.96 \pm 0.33$ kpc is available thanks to Very Long Baseline Interferometry (VLBI) radio parallax observations \citep{Atri_2020}.

Bipolar relativistic discrete ejecta from \maxieight{} have been detected and monitored at radio wavelengths for almost one year, with observations at different angular resolutions with the Multi-Element Radio Linked Interferometer Network (eMERLIN), the Very Long Baseline Array (VLBA), the Arcminute Microkelvin Imager Large Array (AMI-LA), the Karl G. Jansky Very Large Array (VLA) and the MeerKAT radio interferometer, showing the jets to propagate out to $\sim$10 arcsec from the core of the system with a high proper motion \citep{Bright, Wood_2021}. Notably, these jets have also been detected at large scales, up to 12 arcsec, in the X-rays with five \textit{Chandra} X-ray telescope exposures between the end of 2018 and 2019 \citep{Espinasse_xray}. These X-ray detections are particularly important because they cover the deceleration phase, not immediately evident from the radio data alone.
In this work, we use both the radio and X-ray coordinates of the jets to model their motion, and we also take into account the updated jet coordinates from \cite{Wood_2021}, obtained with the application of the new dynamic phase centre tracking technique to the VLBA data (see also \citealt{Wood_2023}).

\subsection{\maxififth{}}
\label{sec:1535}

The BH XRB \maxififth{} was discovered by MAXI in September 2017 \citep{Negoro_2017} when it entered into outburst, and it was subsequently monitored at all wavelengths between radio and the hard X-rays during its 1-year long outburst (e.g.\ \citealt{Tao_2018, Huang_2018, Russell_1535, Parikh, Bhargava_2019, Baglio_2018}). In particular, the full outburst evolution with the associated state transitions is discussed in \cite{Tao_2018} and \cite{Nakahira_2018}. The source is located at a distance $D = {4.1}_{-0.5}^{+0.6}$ kpc \citep{Chauhan_2019}, determined from observations of H\textsc{i} absorption carried out with the Australian Square Kilometre Array Pathfinder (ASKAP).

The radio monitoring campaign presented in \cite{Russell_1535} covered the evolution of the jets from \maxififth{} throughout the whole outburst, with Australia Telescope Compact Array (ATCA) and MeerKAT observations. 
Compact jets were detected during an initial brightening in the first hard state and they were subsequently observed to quench as the source transitioned to the intermediate state, displaying intense flaring activity (see also \citealt{Russell_2020_break_frequency}).
During the hard-to-soft state transition, \maxififth{} launched a fast single-sided discrete jet that was detected and monitored with MeerKAT and ATCA for almost one year. The relativistic components was observed to propagate and decelerate up to an angular distance of $\sim$15 arcsec \citep{Russell_1535}. Its monitoring allowed the authors to place model-independent constraints on the jet speed, inclination angle and ejection date, which we take into account in the modelling presented in this work.

\subsection{\xtej{}}
\label{sec:1752}

\xtej{} is a BH XRB discovered by the Rossi X-ray Timing Explorer in 2009 \citep{Markwardt_2009} that remained active for almost one year in outburst and that has been the subject of dense multi-wavelength observing campaigns, mostly focused in the radio and X-ray bands \citep{Shaposhnikov_2010, Ratti_2012, Brocksopp}. A recent estimation based on the Bayesian analysis of the soft spectral state and the hard-to-soft state transition yields the following constraint on the source distance: $D = 7.11^{+0.27}_{-0.25}$ kpc \citep{Abdulghani_2024}, which is notably more than twice the first distance estimation of $3.5$ kpc from \cite{Shaposhnikov_2010}, and it is consistent with another recent estimation of $D = 6 \pm 2$ kpc based on Gaia DR3 \citep{Fortin_2024}. We note that \cite{Abdulghani_2024} also provide a first BH mass estimation of $12 \pm 1 M_{\odot}$, based on the same method. In this paper, we adopt $7.1$ kpc as the source distance, noting that adopting the $6$ kpc value would not substantially change the main conclusions of this work. 
Also, we do not use the $3.5$ kpc distance, as it is obtained with an uncertain X-ray spectral-timing correlation scaling technique based on the evolution of the photon index and the QPO frequency during the outburst \citep{Shaposhnikov_2009}.

Notably, during the outburst and after the first hard state, the source performed a standard transition to the intermediate and then soft state, but then displayed multiple short-lived returns to the intermediate state accompanied by strong radio flaring activity observed with ATCA, implying the production and launch of multiple ejecta \citep{Brocksopp}. 
These ejecta, at least three separated approaching components, were eventually imaged with the European VLBI Network (EVN) and with VLBA \citep{Yang2010, Yang_2011, MJ_2011}, appearing to propagate only at small scales, i.e. less than one arcsec. At least one component (labeled \enquote{A} in \citealt{Yang2010}) displayed evidence of deceleration \citep{MJ_2011}, while no receding component was detected and no ejecta were detected at larger distances from the core with interferometers probing larger angular scales (such as ATCA). In this paper we will only focus on this decelerating component, as not enough data are available for the other ejecta.

\section{The dynamical model}
\label{sec:The dynamical model}

We adopt a numerical blast-wave dynamical model to describe propagation of jets in the ISM. The model was originally developed to describe GRB afterglows \citep{Piran_1999, Huang_1999}, and has been applied to describe the evolution of mildly relativistic ejecta in BH XRBs for the first time in \cite{Wang_model}, including the transition from relativistic to non-relativistic motion. In particular, we consider the same implementation of \cite{Carotenuto_2022}.

The model considers a symmetric pair of confined conical ejecta launched simultaneously in opposite directions, at an inclination angle $\theta$ with respect to the line of sight. The ejecta start to move away from the core with an initial Lorentz factor $\Gamma_0$ and kinetic energy $E_0$, and expand with a constant half-opening angle $\phi$ inside an ambient medium with constant density $n_{\rm ISM}$. Matter in the same ambient medium is entrained by the jets, which are therefore continuously decelerating, and during this process their kinetic energy is converted into internal energy of the swept-up ISM through external shocks, in a similar fashion as GRB afterglows \citep{Wang_model}. In particular, a forward shock develops at the contact discontinuity between the jet and the ISM. The shock continuously heats the encountered ISM and randomly accelerates particles. In this context, the radiated energy is assumed to be negligible and the jet expansion is considered adiabatic throughout its whole evolution \citep{Chiang_1999, Huang_1999, Wang_model, Hao}, an assumption that has been proven to be robust in recent works \citep{Steiner_xte, Bright, Carotenuto_2022, Zdziarski_2023}.
Considering one of the two ejected components, it is possible to write the energy conservation equation
\begin{equation}
E_0 = (\Gamma -1) M_0 c^2 + \sigma(\Gamma_{\rm sh}^2-1)m_{\rm sw}c^2,
\label{eq:energy}
\end{equation}
where the two terms on the right-hand side are, respectively, the instantaneous kinetic energy of the ejecta and the internal energy of the swept-up ISM. More in detail, $\Gamma$ is instantaneous jet bulk Lorentz factor, $M_0$ is the mass of the ejecta and $\Gamma_{\rm sh}$ is the Lorentz factor of the shocked ISM. Here, $\sigma$ is a numerical factor equal to $6/17$ for ultra-relativistic shocks and $\sim$0.73 for non-relativistic shocks \citep{Blandford_mckee_1976}, and it is possible to interpolate between the two regimes with the following numerical scaling \citep{Huang_1999, Wang_model, Steiner_xte}:
\begin{equation}
\sigma=0.73-0.38\beta,
\label{eq:sigma}
\end{equation}
where $\beta=(1-\Gamma^{-2})^{1/2}$ is the intrinsic jet speed in units of $c$. The mass of the entrained material $m_{\rm sw}$ can be written as
\begin{equation}
m_{ \rm sw} = \frac{\phi^2 \pi m_{\rm p} n R^3}{3},
\label{eq:mass_swept}
\end{equation}
where $R$ is the distance from the compact object, $n$ is the ambient density and $m_{\rm p}$ is the proton mass. The Lorentz factor of the shocked ISM can be expressed as a function of the jet bulk Lorentz factor by using the jump conditions for an arbitrary shock \citep{Blandford_mckee_1976, Steiner_xte}:
\begin{equation}
\Gamma_{\rm sh}^2=\frac{(\Gamma+1)(\hat{\gamma}(\Gamma-1)+1)^2}{\hat{\gamma}(2-\hat{\gamma})(\Gamma-1)+2},
\label{eq:Lorentz_shock}
\end{equation}
where $\hat{\gamma}$ is the adiabatic index that varies between $5/3$ for ultra-relativistic shocks and $4/3$ for non-relativistic shocks. As the jet decelerates, we interpolate between the two limits with
\begin{equation}
\hat{\gamma}=\frac{4\Gamma+1}{3\Gamma}.
\label{eq:interpolation}
\end{equation}

The relativistic kinematic equations for the approaching and receding components are \citep{Rees_1966, Blandford_1977_superluminal, Mirabel1994}:
\begin{equation}
\frac{dR}{dt}=\frac{\beta c}{1 \mp \beta\cos{\theta}},
\label{eq:proper motion}
\end{equation}
where the $\mp$ refers, respectively, to the approaching and receding jet, and $t$ is the arrival time of the photons at the observer.

The measurable projected angular separation from the core is
\begin{equation}
\alpha(t)=\frac{R(t) \sin{\theta}}{D},
\label{eq:alpha}
\end{equation}
where $D$ is the source distance.

In total, the model depends on 7 parameters: the jet initial kinetic energy $E_0$ and Lorentz factor $\Gamma_0$, the inclination angle of the jet axis $\theta$ and the jet half-opening angle $\phi$, the source distance $D$, the external ISM density $n_{\rm ISM}$ and the ejection time $t_{\rm ej}$. It is crucial to note that a degeneracy exists in this model between the three parameters $E_0, \phi$ and $n_{\rm ISM}$, as they appear as a single term in Equation \ref{eq:energy} (taking into account the expression for $m_{ \rm sw}$ in Equation \ref{eq:mass_swept}). Hence, only the factor $E_0/n_{\rm ISM} \phi^2$ can be independently constrained by the application of this model. Similarly to \cite{Steiner_xte}, we refer to such factor as \enquote{effective energy}, which here we define as
\begin{equation}
     \Tilde{E}_0 = E_0 \left( \frac{n_{\rm ISM}}{1 \rm \ cm^{-3}} \right)^{-1} \left(\frac{\phi}{1\degree} \right)^{-2}.
     \label{eq:effective_definition}
\end{equation}
Therefore, in order to obtain a reliable estimate of the jet kinetic energy, one needs to independently measure the two parameters $\phi$ and $n_{\rm ISM}$, or to assume reasonable values for them (see Section \ref{sec:Jet kinetic energy and external ISM density}).

For every set of parameters that compose the model, it is possible to obtain the proper motion curve of the jet on the plane of the sky and then to compare it with the data. This can be done by integrating Equation \ref{eq:proper motion} starting at a time $t_{\rm ej}$ from an assumed distance $R_0 = 10^8$ cm (from which there is only a weak dependence), and then numerically solving Equation \ref{eq:energy} at every time step for the instantaneous jet Lorentz factor. The information on the instantaneous speed is then used to update the distance traveled by the jet, which is converted to the angular separation $\alpha$ (Equation \ref{eq:alpha}), that can be directly compared to the observational data.

\section{Results}
\label{sec:Results}

\subsection{Fit setup}
\label{sec:fit setup}
 
We fit the data for the three BH XRBs considered in this work with the dynamical model presented in Section \ref{sec:The dynamical model}. We adopt a Bayesian approach, applying a Monte Carlo Markov Chain (MCMC) code implemented with the \textsc{emcee} package \citep{emcee_paper}.
For every point of the parameter space, Equation \ref{eq:proper motion} was integrated using {\tt odeint} from the SciPy package \citep{2020SciPy-NMeth}.

We include the maximum amount of available information in the choice of our priors, which are physically motivated from our knowledge of the source in question and of BH XRBs in general. We discuss the specific choices in the following sections dedicated to each source. Every MCMC run was conducted using 110 walkers. For each run, after manual inspection, we consider that convergence is reached when the positions of the walkers in the parameter space are no longer significantly evolving. 
Once the chains have converged, the best fit result for each parameter is taken as the median of the one-dimensional posterior distribution obtained from the converged chains, while the 1$\sigma$ uncertainties are reported as the difference between the median and the 15th percentile of the posterior (lower error bar), and the difference between the 85th percentile and the median (upper error bar).

\subsection{\maxieight{}}
\label{sec:results_1820}

\begin{figure*}
\begin{center}
\includegraphics[width=\textwidth]{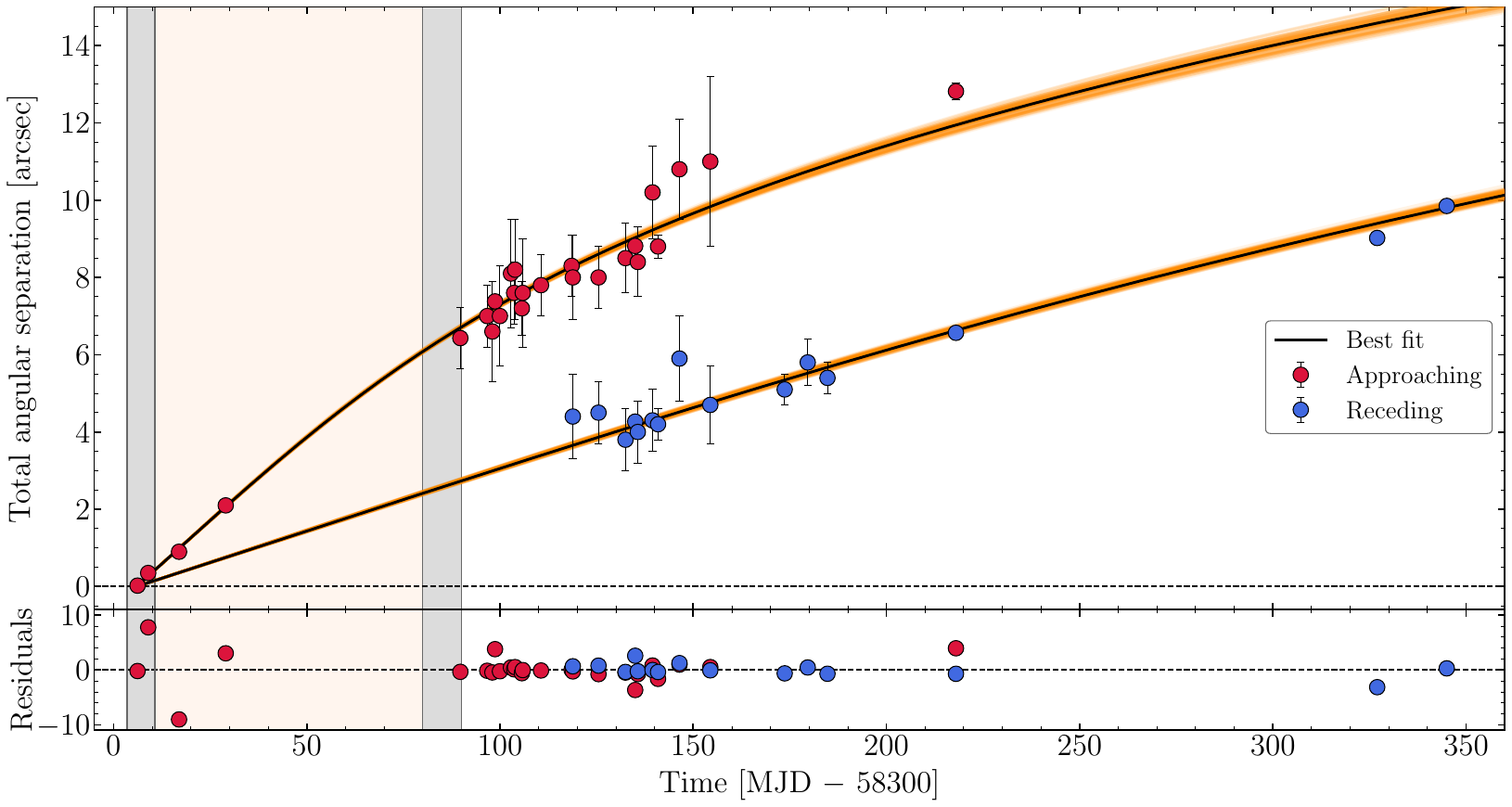}
\caption{Angular separation in arcsec between the discrete ejecta and the position of \maxieight{}, with data from \protect\cite{Bright, Espinasse_xray} and \protect\cite{Wood_2021}. The un-shaded, gray and seashell regions mark periods in which the source was, respectively, in the hard, intermediate and soft state \protect\citep{Shidatsu_2018}. The black horizontal dashed line represents the zero separation from the core, while the black continuous line represents the best fit obtained with the external shock model. The orange shaded area represents the total uncertainty on the fit and it is obtained by plotting the jet trajectories corresponding to the final positions of the MCMC walkers in the model parameter space. Residuals ([data – model]/uncertainties) are reported in the bottom panel. The model appears to provide an excellent description of the motion of both the approaching and receding ejecta, with a low statistical uncertainty.}
\label{fig:angsep_fit_uncertainties_1820}
\end{center}
\end{figure*}

We first consider the bipolar ejecta from \maxieight{}. The angular separation for the two components is shown in Figure \ref{fig:angsep_fit_uncertainties_1820}, including the measurements presented in \cite{Bright}, \cite{Espinasse_xray} and \cite{Wood_2021}. The approaching and receding components are marked, respectively, by red and blue points.
We performed a joint fit of the dynamical model presented in Section \ref{sec:The dynamical model} to the approaching and receding components. We adopted a flat prior for $\Gamma_0$ (between $1$ and $100$) and a log-flat prior for $\Tilde{E}_0$ (between $10^{35}$ and $10^{55}$ erg). We further assumed a normal prior for the source distance centered on $D = 2.96$ kpc and with a width of $0.3$ kpc \citep{Atri_2020}, and we assumed a flat prior for $t_{\rm ej}$ centered on the ejection time of component C (MJD 58305.95) presented in \cite{Wood_2021} and ranging between MJD 58300 and 58310. For the inclination angle, again relying on \cite{Wood_2021}, we used a normal distribution centered on $64\degree$ and with a width of $5\degree$, while truncated outside the interval $0\degree$--$90\degree$. 

The best fit is shown in Figure \ref{fig:angsep_fit_uncertainties_1820}, along with the proper motion of the two jet components, and the results are reported in Table \ref{tab:fit_params_jets}. The statistical uncertainty range on the plot is represented as the ensemble of trajectories corresponding to the final positions of the walkers in the parameter space. From Figure \ref{fig:angsep_fit_uncertainties_1820}, it is possible to see that the model fits exceptionally well to the data, and the agreement with observations can be seen from the residuals on the bottom panel of the same figure. The deceleration of both jets can be adequately described by a single Sedov phase in a homogeneous environment. This type of deceleration has also been modeled using a simple polynomial fit in \cite{Espinasse_xray} and \cite{Wood_2021}, but we note that in our case the whole jet motion can be described by a single physical model. The statistical uncertainty on the fit is remarkably small thanks to the fact that we detected both components and that we had VLBI observations taken at the beginning of the jet motion \citep{Bright, Wood_2021}, which allowed us to constrain with great accuracy the ejection time. The high-resolution Chandra observations at the end of the monitoring are also important to cover the deceleration phase \citep{Espinasse_xray}. According to this model, the jet is launched at $t_{\rm ej} = \rm MJD \ 58305.96^{+0.02}_{-0.02}$ with a bulk Lorentz factor $\Gamma_0 = 2.61^{+0.54}_{-0.39}$, an effective energy of $\Tilde{E}_0 = 2.6^{+0.4}_{-0.4} \times 10^{46} \ {\rm erg}$, and a medium-to-high inclination angle $\theta = 59.6\degree_{-1.0\degree}^{+1.2\degree}$. The source distance is $D = 2.96_{-0.13}^{+0.11}$ kpc, which tracks the prior choice based on the the radio parallax measurement by \cite{Atri_2020}. The posterior distributions for the parameters of the model are shown the Appendix \ref{sec:Posterior distributions} (Figure \ref{fig:corner_1820}), where we present the corner plot displaying the one-dimensional posterior distribution for all the parameters and the two-parameters correlations.

\subsection{\maxififth{}}
\label{sec:results_1535}

\begin{figure*}
\begin{center}
\includegraphics[width=\textwidth]{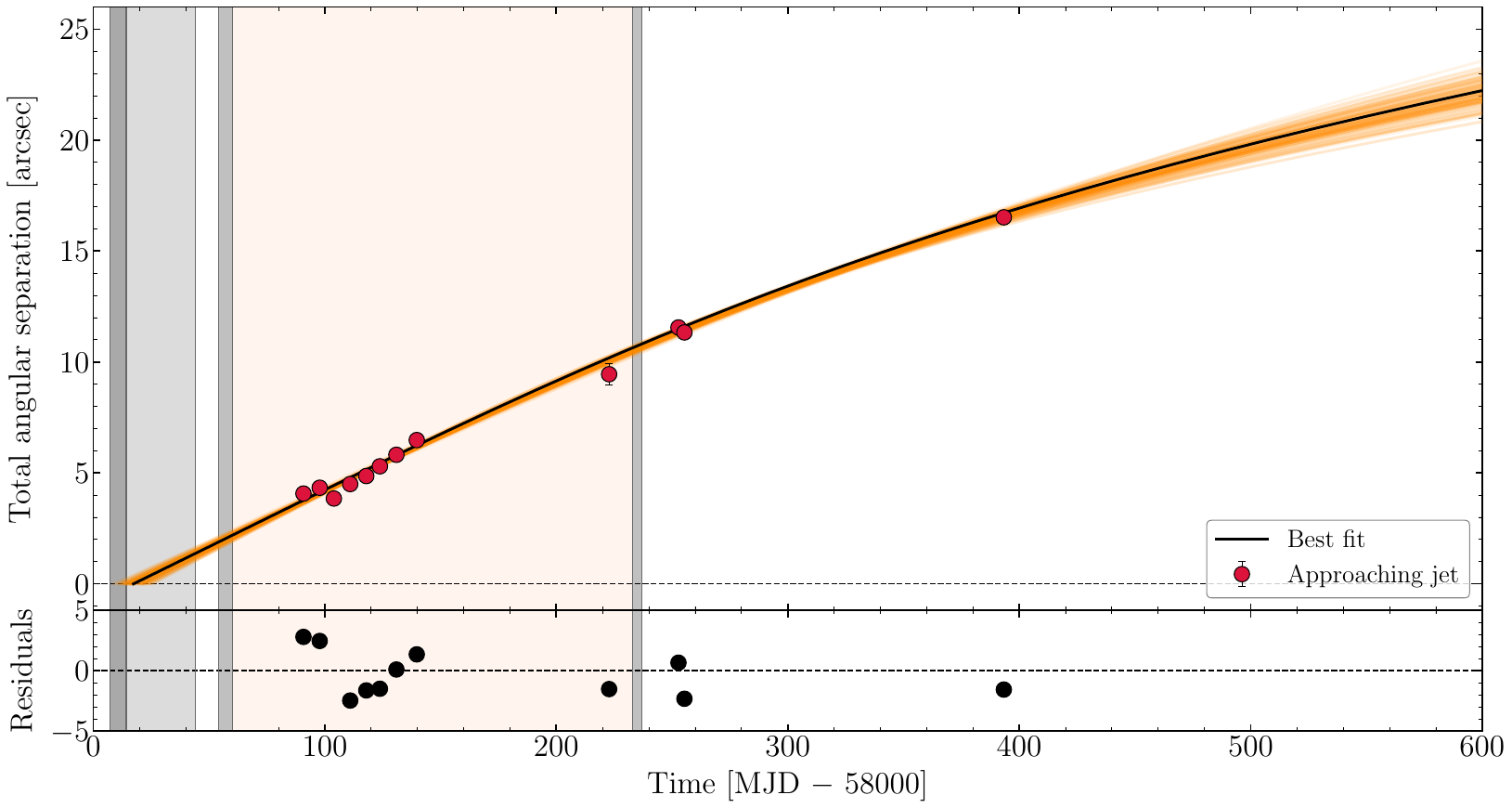}
\caption{Same as Figure \ref{fig:angsep_fit_uncertainties_1820}, but for \maxififth{}, with data from \protect\cite{Russell_1535} and information on the spectral states from \protect\cite{Tao_2018}. Dark and light gray regions differentiate, respectively, between the HIMS and the SIMS. The model appears to fit the data remarkably well, implying that a Sedov phase is an adequate physical scenario for the jet deceleration.}
\label{fig:angsep_fit_uncertainties_1535}
\end{center}
\end{figure*}

We fit the external shock model to large scale jet data from \maxififth{}, using the measurements reported in \cite{Russell_1535}.
Similarly to the fit already presented in \cite{Carotenuto_2022}, we fit the data for the approaching ejection only, given the non-detection of the receding counterpart. The associated proper motion is shown in Figure \ref{fig:angsep_fit_uncertainties_1535}. As for \maxieight{}, we adopted a flat prior for $\Gamma_0$ (between $1$ and $100$) and log-flat prior for $\Tilde{E}_0$ (between $10^{35}$ and $10^{55}$ erg). We assumed a flat prior for $t_{\rm ej}$ between MJD 58005 and 58025 and a normal prior for the source distance centered on $D = 4.1$ kpc, with a width of $0.5$ kpc \citep{Chauhan_2019}. For the inclination angle, we used a uniform distribution in $\cos{\theta}$ truncated outside the interval $0\degree$--$45\degree$, following the constraints reported in \cite{Russell_1535}. 

The best fit results are reported in Table \ref{tab:fit_params_jets} and are shown in Figure \ref{fig:angsep_fit_uncertainties_1535}, from which can be seen that the model fits the data remarkably well. The bottom panel of the figure displays the residuals, revealing a good agreement with the observations. Also in this case, the jet deceleration can be accurately described by a single Sedov phase in a homogeneous environment. From the fit, we can place constraints on the jet ejection date $t_{\rm ej} = \rm MJD \ 58017.4^{+4.0}_{-3.8}$, on its bulk Lorentz factor $\Gamma_0 = 1.6^{+0.2}_{-0.2}$ and on its medium-to-low inclination angle $\theta =30.3\degree^{+6.3\degree}_{-6.3\degree}$. The source distance is $D = 4.2^{+0.8}_{-0.9}$ kpc, which, also in this case, tracks the prior choice based on the the H\textsc{i} absorption measurement by \cite{Chauhan_2019}. We also constrain the effective energy of the jets to be $E_0 = 5.8^{+16.6}_{-4.0} \times 10^{48} \ {\rm \rm erg}$. The posterior distributions for the parameters of the model are shown in Figure \ref{fig:corner_1535}.

\subsection{\xtej{}}
\label{sec:Results_xte}

\begin{figure*}
\begin{center}
\includegraphics[width=\textwidth]{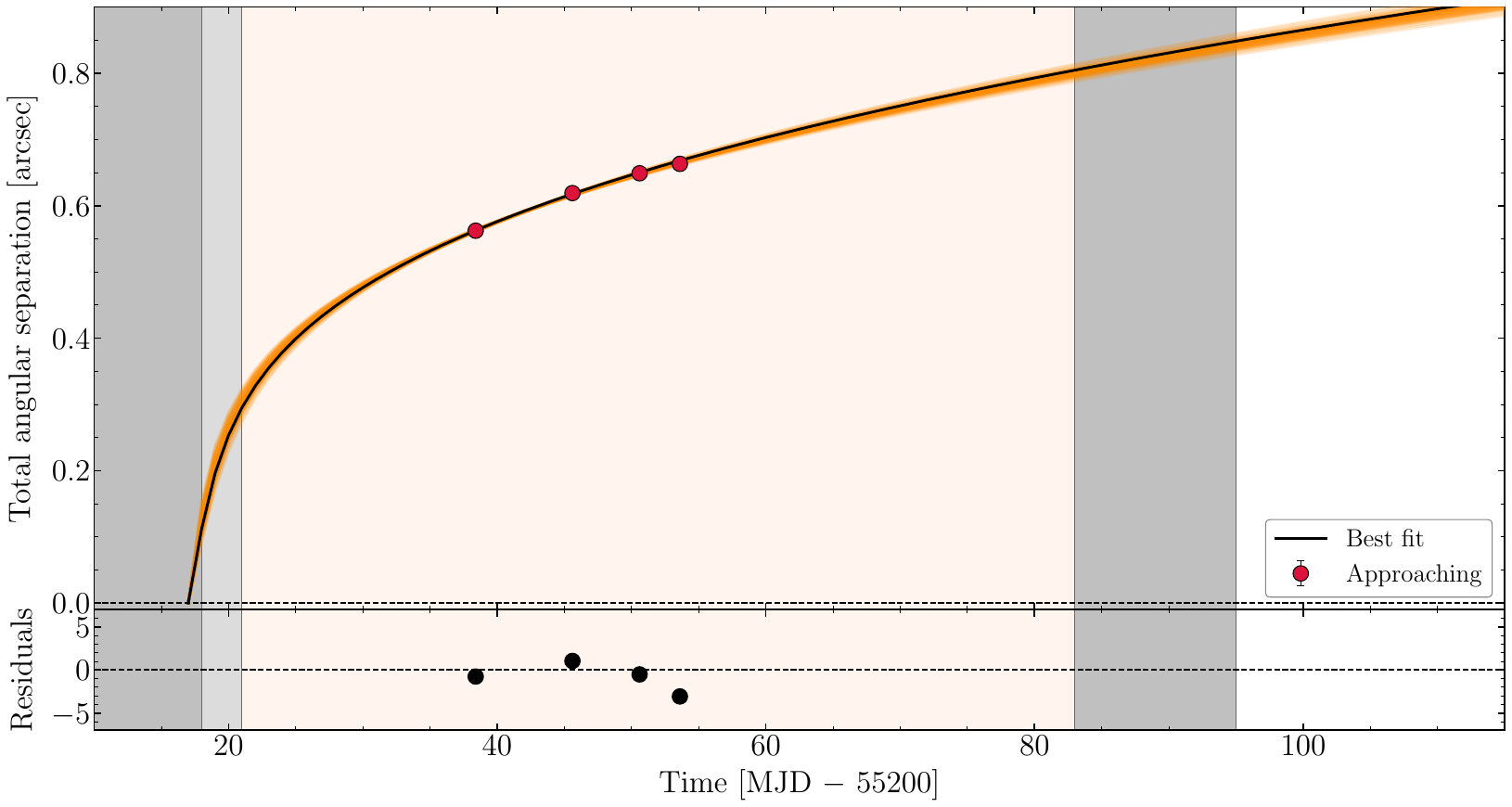}
\caption{Same as Figure \ref{fig:angsep_fit_uncertainties_1535}, but for \xtej{}, with data from \protect\cite{MJ_2011} and information on the spectral states from \protect\cite{Shaposhnikov_2010} and \protect\cite{Brocksopp}. Once the ejection date is fixed (here at MJD 55217), the model with a single Sedov phase appears to fit reasonably well the data.}
\label{fig:angsep_fit_uncertainties_1752}
\end{center}
\end{figure*}

Finally, we fit the dynamical model to the VLBI data of the approaching ejecta launched by \xtej{}, as reported in \cite{MJ_2011}. As for the previous cases, we adopted a flat prior for $\Gamma_0$ (between $1$ and $100$) and a log-flat prior for $\Tilde{E}_0$ (between $10^{35}$ and $10^{55}$ erg). We assumed a uniform distribution in $\cos{\theta}$, with a truncation outside the interval 0\degree--45\degree, consistent with the constraints reported in \cite{MJ_2011}. Moreover, we assume a normal prior for the source distance centered on $D = 7.1$ kpc and with a width of $0.3$ kpc, from \cite{Abdulghani_2024}. Given that only four data points are available, we fixed the ejection date $t_{\rm ej}$ to MJD 55217, just before the $20$ mJy peak of the radio flare observed with ATCA \citep{Brocksopp}, and one day before the transition from the hard-intermediate state (HIMS) to the soft-intermediate state (SIMS) occurred around MJD 55218 \citep{Shaposhnikov_2010}.

The best fit results are reported in Table \ref{tab:fit_params_jets} and are shown in Figure \ref{fig:angsep_fit_uncertainties_1752}, along with the proper motion of the jet, while the posterior distributions for the model parameters are shown in Figure \ref{fig:corner_1752}. According to the model, the jet is launched with an effective energy of $\Tilde{E_0} = 1.1^{+1.2}_{-0.6} \times 10^{45} \ {\rm \rm erg}$, at a medium-to-low inclination angle of $\theta =18.4\degree_{-2.3\degree}^{+2.5\degree}$. Given the scarcity of data on this source, we are unable to provide a robust value for the jet initial Lorentz factor $\Gamma_0$, but from the posterior we can constrain $\Gamma_0 > 3.4$ (99.7\% of confidence), according to the choice of ejection date discussed above. As shown in Figure \ref{fig:corner_1752}, the data provide a median value of $\simeq$5.4, but this constraint depends directly on the chosen $t_{\rm ej}$. In the same way, the lower limit on $\Gamma_0$ can be relaxed if we assume an earlier ejection date. For illustration, performing the same fit, but assuming $t_{\rm ej} = \rm MJD \ 55210$ or $55215$ leads to $\Gamma_0 > 1.8$ (median value $\simeq 2.1$) and $\Gamma_0 > 2.9$ (median value $\simeq 4.2$), respectively.  
On the other hand, for ejections at later times, fixing $t_{\rm ej} = \rm MJD \ 55218$ at the transition from the HIMS to the SIMS \citep{Shaposhnikov_2010} results in an equally acceptable fit, yielding similar constraints for the initial jet Lorentz factor: $\Gamma_0 > 3.4$ (99.7\% of confidence, but with a median value $\simeq 7.4$). Similarly, fixing $t_{\rm ej} = \rm MJD \ 55221$ at the transition from the SIMS to the soft state (an ejection time which we deem more unlikely for BH XRBs) yields $\Gamma_0 > 3.5$ (99.7\% of confidence), with an extremely high median value of $\simeq 13.4$. We note that, for our initial Lorentz factor posterior distributions, fixing $t_{\rm ej}$ in the range MJD 55217-55821 results in very similar lower limits on $\Gamma_0$, while the median value is much more sensitive to the choice of $t_{\rm ej}$.

\setlength{\tabcolsep}{8pt}
\setlength{\extrarowheight}{.7em}
\begin{table*}
\caption{Parameters of the blast-wave model applied in this work inferred from the Bayesian fit described in Section \ref{sec:Results}. The values quoted are the median parameter and the $1\sigma$ confidence intervals derived from the MCMC run. The effective energy is defined as $\Tilde{E}_0 = E_0/n_{\rm ISM}\phi^2$ (see Section \ref{sec:The dynamical model}, Equation \ref{eq:effective_definition}).\\ $^\dagger$: The constraint on $\Gamma_0$ for \xtej{} strongly depends on the preferred ejection date (see Section \ref{sec:Results_xte}).}
\label{tab:fit_params_jets}
\begin{tabular}{*{5}{c l c c c}}
\hline
\hline
Parameter        & Description 							           &    MAXI J1820$+$070         &  MAXI~J1535--571          &  XTE~J1752--223\\
\hline
                                    
$\Gamma_0$       & Initial bulk Lorentz factor at launch           &   $2.6_{-0.4}^{+0.5}$     & $1.6_{-0.2}^{+0.2}$   & $> 3.4^\dagger$\\
$\Tilde{E}_0$ & Effective energy (erg)          &   $2.6^{+0.4}_{-0.4} \times 10^{46}$    & $5.8^{+16.6}_{-4.0} \times 10^{48}$  & $1.1^{+1.2}_{-0.6} \times 10^{45}$\\
$\theta$         & Inclination angle (\degree) 			           &   $59.6_{-1.2}^{+1.0}$       & $30.3_{-6.3}^{+6.3}$  & $18.4_{-2.3}^{+2.5}$\\
$D$     & Source distance (kpc) 		           				   &   $2.96_{-0.13}^{+0.11}$ & $4.2_{-0.9}^{+0.8}$  & $7.0_{-0.1}^{+0.1}$\\
$t_{\rm ej}$     & Ejection date (MJD) 		           				   &   $58305.96_{-0.02}^{+0.02}$ & $58017.4_{-3.8}^{+4.0}$  & $55217 \ (\rm fixed)$\\   
\hline                                                                                                                                                                                                                                                                                                                                                        \end{tabular}
\end{table*}

\section{Discussion}
\label{sec:Discussion}

We successfully modeled the motion of the large-scale jets from three BH XRBs, \maxieight{}, \maxififth{} and \xtej{} with a dynamical blast-wave model based on external shocks. This physical model is found to provide an excellent description of the propagation of the ejecta in the ISM, regardless of whether we detect or not both ejected components, and it also allows us to place meaningful constraints on various parameters of the ejecta.
After the application of this model to the jets of \xte{}, \hh{} and \maxithirt{} \citep{Steiner_xte, Steiner_h17, Carotenuto_2022}, and considering the results shown in the previous section, it appears that all the large-scale jets display a deceleration consistent with a Sedov phase. 
The goodness of the fits obtained for all the sources confirm the validity of the application of physical models derived from our knowledge of GRBs to XRBs (e.g.\ \citealt{Wang_model}), and highlights the potential of using the largely developed set of theoretical GRB models (including their entire multi-wavelength emission) to explain even more features observed in the less-relativistic jets from XRBs, as for instance the presence of forward and reverse shocks within the jet. We discuss in the following sections the constraints on the jet physical parameters and on the source environment that we obtained in this work, comparing them with our current knowledge of jets from BH XRBs.

\subsection{Lorentz factor}
\label{sec:Lorentz factor}

We first discuss the constraints on the initial Lorentz factor $\Gamma_0$ for the ejecta in our sample. It is generally difficult to constrain this parameter from the simple observation of the jet propagation, especially if the ejecta are significantly superluminal. The reason is that a source of significantly relativistic jets (with proper motions $\mu_{\rm app}$ and $\mu_{\rm rec}$) will usually be observed close to its maximum allowed distance $D_{\rm max} = c (\mu_{\rm app} \mu_{\rm rec})^{-1/2}$, where $\Gamma_0$ diverges \citep{Fender1999, Fender_2003}. Therefore, only lower limits on $\Gamma_0$ are available for most of the sources displaying discrete ejecta (e.g.\ \citealt{Fender_2003, Miller-jones2006, Bright}).

For \maxieight{}, we obtain an interesting estimate of the initial Lorentz factor of the jets: $\Gamma_0 = 2.6_{-0.4}^{+0.5}$, which implies a mildly relativistic ejecta. Such constraint is consistent from the previous lower limit $\Gamma_0 > 2.1$ \citep{Bright, Wood_2021}, and, interestingly, this determination has been possible despite the fact that \maxieight{} is located close to its $D_{\rm max} = 3.1$ kpc \citep{Wood_2021}.

In the case of \maxififth{}, we are also able to place an important constraint on the initial Lorentz factor of the approaching component $\Gamma_0 = 1.6_{-0.2}^{+0.2}$, implying a relatively slow ejecta, traveling initially at $\sim$0.77$c$. Jets from \maxififth{} appear to be among the less relativistic in the observed sample of discrete ejecta \citep{Miller-jones2006, Steiner_xte, Steiner_h17, Carotenuto_2022}.
Interestingly, the ejecta from \maxififth{} is the one that propagates to the largest distance from the core (up to 0.8 pc). Launched with a high $\Tilde{E}_0$ (see Table \ref{tab:fit_params_jets}), it is likely among the most energetic and the least relativistic jets observed so far, similar in nature to \maxithirt{}, which also displayed large scale ejecta propagating up to 0.6 pc from the core, with $\Gamma_0 = 1.85_{-0.12}^{+0.15}$ \citep{Carotenuto2021, Carotenuto_2022}. Since it appears that jets from \maxithirt{} and \maxififth{} are among the most energetic observed so far, this likely implies that the mass content of the ejecta is probably the driving factor that determines the large distance at which the ejecta propagate, with $M_0$ being the dominant factor in the $(\Gamma_0 -1) M_0 c^2$ kinetic energy term in Equation \ref{eq:energy} (see also the discussion in \citealt{Zdziarski_2024}). At the same time, again similarly to \maxithirt{}, such determination of $\Gamma_0$ is one of the most precise and robust (although model-dependent) constraints on the Lorentz factor of a jet from a BH XRB to-date, with this being due to the ideal combination of low $\Gamma_0$ and low inclination angle, which is less affected by the degeneracy between $\theta$ and the source distance (\citealt{Fender_2003}, Fender et al., \textit{in prep.}).




We can only provide a lower limit $\Gamma_0 > 3.4 $ for the ejecta of \xtej{}, which, however, directly depends on the chosen $t_{\rm ej}$, and it can be relaxed if we assume an earlier ejection date, as mentioned in Section \ref{sec:1752}. On the other hand, if $t_{\rm ej}$ is fixed to closer to the state transition, our robust lower limit on $\Gamma_0$ does not vary (see Section \ref{sec:Results_xte}). However, if the ejecta were launched on MJD 55218, at the peak of the radio flare on the day of the HIMS-to-SIMS transition, they would have a most likely $\Gamma_0 \simeq 7.4$, which would be the highest Lorentz factor ever observed for these objects and it would challenge the common assumption that jets from BH XRBs are only mildly relativistic, unlike what is observed in AGN and GRBs (e.g.\ \citealt{Jorstad_2005, Ghirlanda_2018}). Observations of the ejecta closer to the core would have greatly helped to determine the initial Lorentz factor of this source, which is likely to be higher than the average among the available sample of ejecta. Overall, the data available so far appear to suggest that BH XRBs are able to accelerate relativistic jets in the mildly relativistic range 
$1 \lesssim \Gamma_0 \lesssim 2$, but a growing number of possible exceptions that suggest even faster jets (as for instance \maxieight{} and \xtej{}), and this appears to be similar to the range of generally inferred bulk Lorentz factor for compact jets (e.g. \citealt{Casella_2010, Saikia, Tetarenko_2019, Tetarenko_2021, Zdziarski_2022}).

\subsection{Inclination angle}
\label{sec:Inclination angle}

The inclination angle that we obtain for \maxieight{} is $\theta = 59.6\degree^{+1.0\degree}_{-1.2\degree}$.
We note that our posterior is still consistent with the chosen normal prior distribution $\theta = 64\degree \pm 5\degree$ from \cite{Wood_2021}. However, the peak of the posterior is slightly lower than the peak of the prior, which is based on the first part of the jet motion.
With the values obtained for $\Gamma_0$ and $\theta$, we can compute the Doppler factor for the two ejecta. The Doppler factor for the approaching component at launch is $\delta_{\rm app} = \Gamma_0^{-1}(1 - \beta_0 \cos{\theta})^{-1} \simeq 0.7$, while for the receding is $\delta_{\rm rec} = \Gamma_0^{-1}(1 + \beta_0 \cos{\theta})^{-1} \simeq 0.25$. This implies that at the beginning of their motion both components are Doppler de-boosted, and their intrinsic luminosity is decreased, respectively by a factor $\delta_{\rm app}^{3-\alpha} \simeq 0.3$ and $\delta_{\rm rec}^{3-\alpha} \simeq 0.008$, using the formalism for discrete jet components, where the flux density follows $S_{\nu} \sim \nu^{\alpha}$ (e.g.\ \citealt{Blandford_1977_superluminal}) and a spectral index $\alpha = -0.6$ \citep{Espinasse_xray}.

We constrain the inclination angle of the jet axis in \maxififth{} to be $\theta = 30.3\degree^{+6.3\degree}_{-6.3\degree}$. Already, from the simple measurement of the jet proper motion and the source distance, \cite{Russell_1535} strongly constrained the maximum jet inclination to $\theta < 45\degree$ (which is included in our choice of the prior on $\theta$).  Notably, such a value does not appear to be consistent with the inclination angle of the inner edge of the accretion disk obtained from NICER X-ray observations \cite{Miller_2018}. The authors report an angle $i = 67.4\degree \pm 0.8\degree$ from the spectral fitting of the relativistic reflection component in the high-SNR X-ray spectrum obtained during the intermediate state (they also report a near maximal BH spin of $a = 0.994 \pm 0.002$, \citealt{Miller_2018}). At the same time, our value is broadly consistent with the inclination angle ($i = 37\degree^{+22\degree}_{-13\degree}$) of the region emitting a narrow and asymmetric iron line \citep{Miller_2018}. Interestingly, the authors explain such difference in $i$ with potential disk warping. The results of this comparison appear counter-intuitive, as we would generally expect jets to be launched along the direction of the BH spin, which, in turn, should be aligned with the inner edge of the accretion disk \citep{Bardeen_Petterson}. However, at least a case in which jets were launched along the axis of a rapidly precessing inner disk has been observed \citep{Miller-Jones2019}. As in the case of \maxithirt{} \citep{Carotenuto_2022}, a measurement of the orbital plane of the system would be useful to test the potential 3D alignment between the disk and the jet axis, taking into account that cases of disk/jet misalignment have been observed \citep{Miller-Jones2019, Poutanen_2022} and this is also supported by GRMHD simulations (e.g.\ \citealt{Liska_2018}). The Doppler factor for the approaching component at launch is $\delta_{\rm app} \simeq 1.9$, implying a Doppler boosting with a factor $\delta_{\rm app}^{3-\alpha} \simeq 10$, while for the receding component we can compute $\delta_{\rm rec} = \simeq 0.37$, implying a de-boosting of a factor $\delta_{\rm rec}^{3-\alpha} \simeq 0.03$, again adopting $\alpha = -0.6$. Such a high Doppler de-boosting easily explains the non-detection of the receding component, as the radio flux density was pushed below the ATCA and MeerKAT sensitivity limit for the available exposure times \citep{Russell_1535}.

Lastly, we infer a low inclination angle of $\theta = 18.4\degree^{+2.5\degree}_{-2.3\degree}$ for \xtej{}. The low inclination angle, coupled to the deceleration at relatively small scales ($<1$ arcsec), can explain why these jets were not resolved with ATCA at larger scales \citep{Brocksopp}. To compute the Doppler factors for the approaching and receding components, we assume a most likely value $\Gamma_0 = 3.5$ (Section \ref{sec:Results_xte}). For the two components, we obtain, respectively, $\delta_{\rm app} \simeq 3.2$ (boosting factor $\delta_{\rm app}^{3-\alpha} \simeq 60$) and $\delta_{\rm rec} \simeq 0.1$ (boosting factor $\delta_{\rm rec}^{3-\alpha} \simeq 10^{-3}$), which is again consistent with the non-detection of the receding component \citep{Yang2010, Yang_2011, MJ_2011}.

\subsection{Ejection time}
\label{sec:Ejection time}

The ejection time is a crucial piece of the puzzle in the current effort to reconstruct the precise sequence of events that lead to the formation and launch of discrete ejecta, which is still unclear, and modelling the jet motion is a reliable way of obtaining such information.

\setlength{\extrarowheight}{.2em}
\setlength{\tabcolsep}{10pt}
\begin{table}
\caption{Time delay $\Delta t_{\rm ej, X}$ between the inferred ejection times $t_{\rm ej}$ and the possibly associated observed X-ray signatures. Here a positive $\Delta t_{\rm ej, X}$ means that the the jet is launched before the first appearance of the X-ray signature. The uncertainties on $\Delta t_{\rm ej, X}$ are the same as the ones obtained for $t_{\rm ej}$ in this work and in \protect\cite{Carotenuto_2022}. For \maxieight{}, we include both Components A (slow jet) and C (fast jet) as labeled in \protect\cite{Wood_2021}, reminding the reader that in this work we only focus on Component C. The uncertainty on \maxififth{} is due to the large uncertainty on $t_{\rm ej}$ and to the fact that Type-B QPOs might have been present between MJD 58016.8 and 58025 \protect\citep{Stevens_2018}.}
\centering
\begin{tabular}{l l l}
\hline
\hline
Source & X-ray signature & $\Delta t_{\rm ej, X}$ [d]\\
\hline
\maxieight{}/A & Type-B QPO & $\sim 0.08^{+0.04}_{-0.04}$\\
\maxieight{}/C & Type-B QPO & $\sim -0.25$\\
\maxififth{} & Type-B QPO & $\sim -0.6^{+12.0}_{-4.0}$\\
\xtej{} & Type-B QPO & $> 0$\\
\maxithirt{} & Type-B QPO & $\sim1^{+3}_{-2}$\\
\hline
\end{tabular}
\label{tab:qpo}
\end{table}

In the case of \maxieight{}, we infer an ejection date $t_{\rm ej} = \rm MJD \ 58305.96^{+0.02}_{-0.02}$, which is completely consistent with the most up-to-date estimation reported in \cite{Wood_2021}. We note that the quality of the data from \cite{Bright}, especially the dense VLBI monitoring at the hard-to-soft state transition, and the additional significant improvements obtained with the new dynamic phase-center tracking technique adopted in \cite{Wood_2021}, already allowed the authors to constrain the ejection time with great accuracy, and it is worth noting that a precision of roughly 30 minutes has rarely been achieved for this type of events. It is also worth nothing that the obtained $t_{\rm ej}$ places the jet launch close to the peak of the radio flare observed at 15.5 GHz with AMI-LA \citep{Bright, Homan_qpo}. Moreover, this jet appeared to be launched approximately 6 hours after the first detection of Type-B QPOs in this outburst, defining the beginning of the soft-intermediate state \citep{Homan_qpo}. However, it is important to remark that \cite{Wood_2021} associate the radio flare and the detection of Type-B QPOs with the ejection of a different pair of ejecta (with the approaching jet labeled as Component A), which had an intrinsic speed $\beta \sim 0.3$, were not detected beyond the  milli-arcsec scale and were ejected $\sim$9 h before the fast large-scale jets that we consider in this work \citep{Wood_2021}. Specifically, Component A displayed an elongated structure and its ejection was inferred to last $\sim$6 h, hence partially overlapping with the first appearance of Type-B QPOs in this source. Interestingly, \cite{Wood_2021} do not identify any X-ray or radio flare counterpart to Component C. It has been suggested that Type-B QPOs could correspond to the time of jet launching (e.g.\ \citealt{Fender_2009, Miller-Jones_h1743}), implying a strong causal relation between the two phenomena that has been particularly highlighted in \cite{Homan_qpo}. However, for \maxieight{} such link appears to be stronger with Component A than with Component C. Therefore, if this timing signature is indeed linked to jet ejections, it is unclear at the moment whether there is any connection with the fast ejecta observed to propagate at large scales. Such connection is also not confirmed for other sources \citep{Miller-Jones_h1743, Russell_1535, Carotenuto2021}, for which in general the ejections are inferred to happen from hours to days before the detection of Type-B QPOs, as can also be seen in Table \ref{tab:qpo} (for which we include data from \maxithirt{},  \citealt{Carotenuto_2022}). 
 

Regarding \maxififth{}, we infer $t_{\rm ej} = \rm MJD \ 58017.4^{+4.0}_{-3.8}$, with a much larger uncertainty with respect to \maxieight{}, mostly due to the lack of early-time VLBI observations of the ejecta. Interestingly, our new $t_{\rm ej}$ places the ejection in the soft intermediate state \citep{Tao_2018}, updating the previous estimation in the hard-intermediate state \citep{Russell_1535}. Given the large radio flare reported on MJD 58017.4 by \cite{Russell_1535}, it is more likely that the ejecta was launched before this date than later, despite our statistical uncertainty being almost symmetrical around the same MJD 58017.4.
Interestingly, our preferred ejection date is now roughly 4 days after the well-monitored quenching of the compact jets, which, from the tracking of the evolution of the break frequency from the infrared to the radio bands, appeared to be switched off over a timescale of 1 d on MJD 58013 \citep{Russell_2020_break_frequency}. If such a result is confirmed, it would imply that discrete ejecta do not result immediately from the destruction of the compact jets, but that instead they are formed and launched sometime afterwards (see also \citealt{Echiburu_2024} for \maxieight{}). 

For \maxififth{}, a tentative detection of Type-B QPOs with NICER was reported in \cite{Stevens_2018}. Specifically, possible Type-B QPOs were detected when stacking the NICER data in the range between MJD 58016.8 and 58025, but it is worth remarking that the authors could not clearly differentiate between Type-A and Type-B QPOs in the data \citep{Stevens_2018}. The reported QPO interval overlaps with our inferred ejection date $t_{\rm ej} = \rm MJD \ 58017.4^{+4.0}_{-3.8}$, but, given the 4-d uncertainty, we cannot precisely conclude on the exact sequence of events. However, if Type-B (or Type-A) QPOs were present during the whole stacked interval and lasted until MJD 58025, then we could at least conclude that they persist after the ejecta are produced.


Unfortunately, we are unable to provide constraints on the ejection date for \xtej{}, given that $t_{\rm ej}$ is a fixed parameter in our modelling. This is also due to the fact that, considering the four available detections, the jets were already strongly decelerating \citep{Yang2010, MJ_2011}. The radio flare peaking peaking on MJD $55218$ and reported in \cite{Brocksopp} suggests that the ejection might have happened on that day or before, while it is unlikely to have happened at later times. Our choice of $t_{\rm ej}$ on MJD 55217 (one of the possible options) places the ejection in the hard-intermediate state. Type-B QPOs were instead reported on MJD 55218 and 55220 \citep{Shaposhnikov_2010}, possibly again suggesting that these variability features could persist after the launch of discrete ejecta. However, the data on \xtej{} do not allow us to draw any firm conclusion on this particular aspect of the jet production.


A summary of the time delay $\Delta t_{\rm ej, X}$ between the jet ejection date and the first appearance of Type-B QPOs for the sources considered in this work is reported in Table \ref{tab:qpo}. We mention that, rather than the appearance/disappearance of QPOs, it has been recently proposed that jet ejections in \grs{} could be linked to a change in the coronal geometry observed through changes in the Type-C QPO frequency and a change of sign in the phase lags at the QPO frequency, along with the simultaneous radio emission \citep{Mendez_2022}. This result (see also \citealt{Garcia_2022}) appears to be also consistent with previous findings on the non-trivial geometry of the corona in \maxieight{} \citep{Kara_2019} and \maxithirt{} \citep{Garcia_2021}.

\subsection{Jet kinetic energy and external ISM density}
\label{sec:Jet kinetic energy and external ISM density}

We discuss in this section our results for the jet kinetic energy. Due to the degeneracy between $E_0$, $\phi$ and $n_{\rm ISM}$ (see Section \ref{sec:The dynamical model}), only the effective energy $\Tilde{E}_0 = E_0/n_{\rm ISM}\phi^2$ can be independently constrained by our fit. This parameter can be easily interpreted as the kinetic energy required for a $\phi = 1\degree$ jet to propagate to the observed distance in a $1$ cm$^{-3}$ ISM. Considering our sample, we constrain an effective energy of $\Tilde{E}_0 = 2.6^{+0.4}_{-0.4} \times 10^{46} \ {\rm erg}$ for \maxieight{}, a higher value $\Tilde{E}_0 = 5.8^{+16.6}_{-4.0} \times 10^{48} \ {\rm \rm erg}$ for \maxififth{} and a lower value $\Tilde{E}_0 = 1.1^{+1.2}_{-0.6} \times 10^{45} \ {\rm \rm erg}$ for \xtej{}. The highest value obtained for \maxififth{} can be explained by the fact that, assuming the same $n_{\rm ISM}$, the jets from \maxififth{} propagate up to a larger angular distance with less evident deceleration when compared to the other two sources.

\begin{figure*}
\begin{center}
\includegraphics[width=8cm]{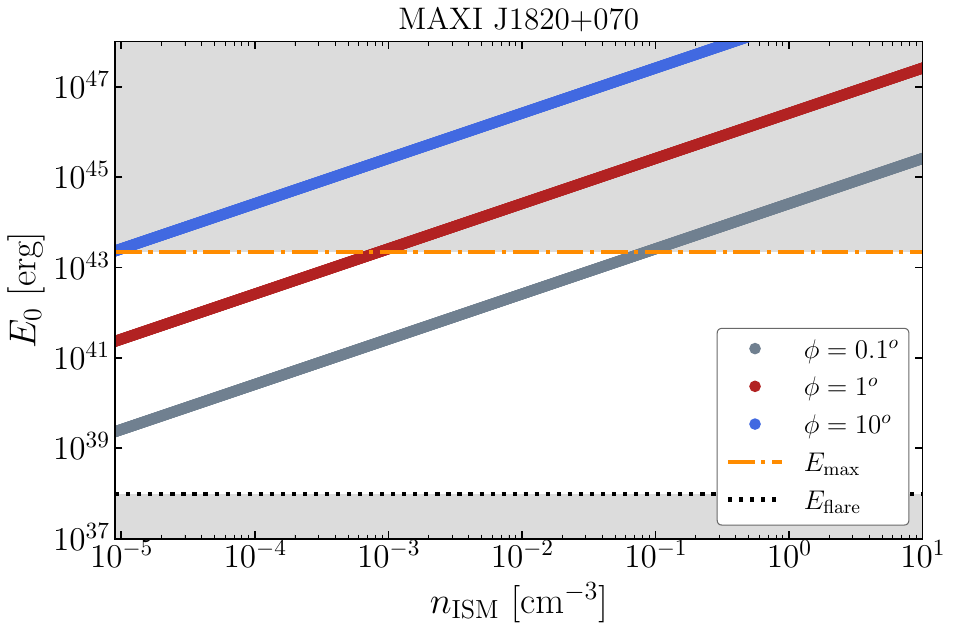}\includegraphics[width=8cm]{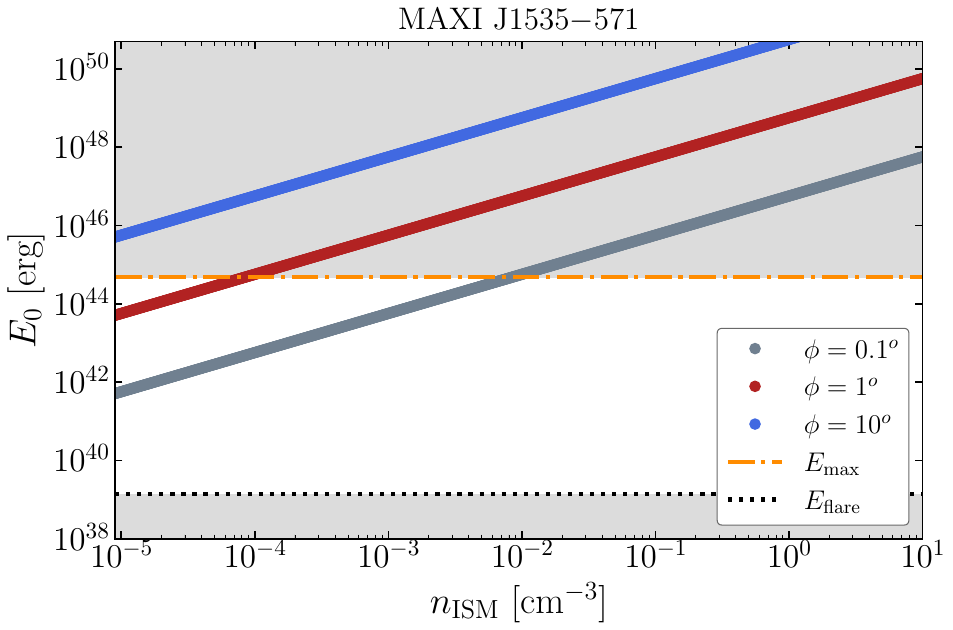}\\ \includegraphics[width=8cm]{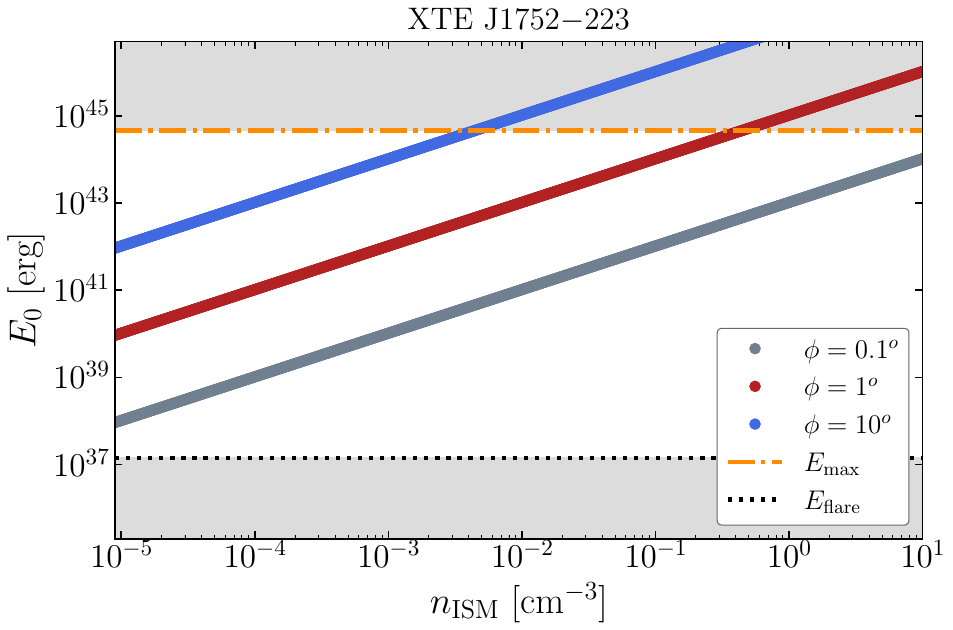}\includegraphics[width=8cm]{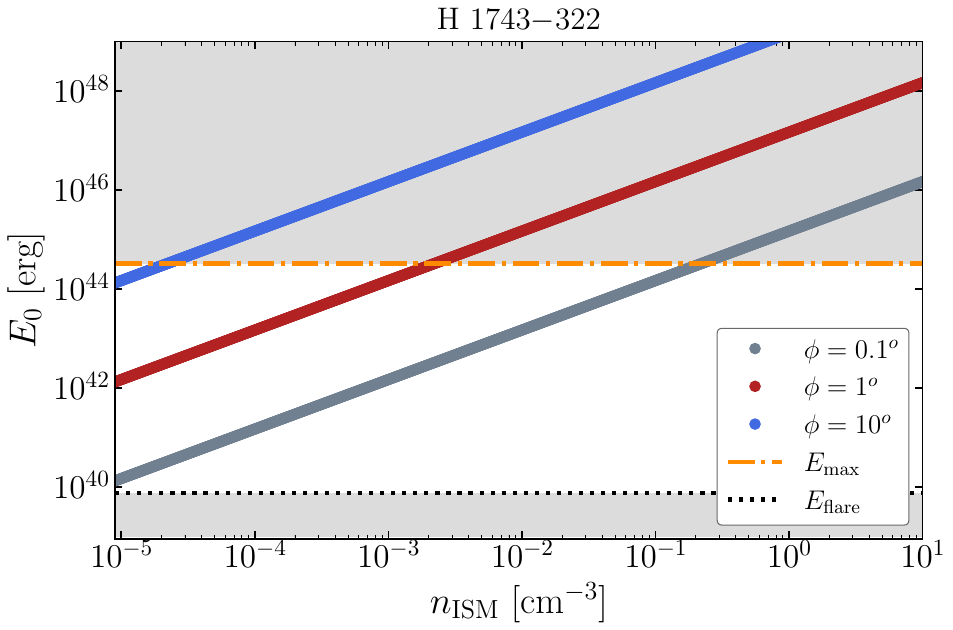}
\caption{Explicit dependence of the jet kinetic energy $E_0$ on the external ISM density $n_{\rm ISM}$ for different values of the half-opening angle $\phi$. The dependence is obtained through the constraints on the effective energy $\Tilde{E}_0$ and it is here shown for the four sources considered in Section \ref{sec:Jet kinetic energy and external ISM density}. The horizontal dot-dashed orange line represents the maximum energy $E_{\rm max}$ available to the jet from the simultaneous accretion power, which sets a strong upper limit on $n_{\rm ISM}$ for all sources except \xtej{}. The dotted black line shows instead the minimum energy $E_{\rm flare}$ in the jet frame inferred from the radio flare associated to the ejection, here implying that such value is likely a large underestimation of the jet kinetic energy, given that it would require extremely low value of $n_{\rm ISM}$ for the jet to propagate up to the observed distances. Regions excluded by our constraints on $E_{\rm max}$ and $E_{\rm flare}$ are shaded in grey.}
\label{fig:E0_vs_nism}
\end{center}
\end{figure*}

Through the constraints on $\Tilde{E}_0$, we plot in the first three panels of Figure \ref{fig:E0_vs_nism} the explicit dependence of $E_0$ on $n_{\rm ISM}$ for different values of the jet opening angle. Given that no independent information is available regarding the exact values of $n_{\rm ISM}$ and $\phi$ for any of our sources, it is not possible to provide a preferred estimate of the jet kinetic energy. Regarding the jet half-opening angle, the value $\phi = 1\degree$ is generally adopted in the literature \citep{Steiner_xte, Steiner_h17, Carotenuto_2022}. Such value is consistent with the large number of observational upper limits available for these jets (e.g.\ \citealt{Kaaret_2003, Miller-jones2006, Russell_1535, Carotenuto2021, Wood_2021, Williams_2022}). Moreover, for cases in which these jets have been resolved in radio or X-rays, the inferred opening angles were of the same order of magnitude (e.g.\ \citealt{Bright, Espinasse_xray, Chauhan_2021_1535}). Therefore, we assume $\phi = 1\degree$ as a reasonable choice, but we also show our solutions for narrower or less-likely wider jets in Figure \ref{fig:E0_vs_nism}.

On the other hand, much less information is available on the density of the environment of BH XRBs. Two other BH XRBs that displayed large-scale decelerating jets, \xte{} and \maxithirt{}, were inferred to be located in a low-density ISM cavity, for which an reasonable external ISM density of $1$ 
cm$^{-3}$ was assumed \citep{Steiner_xte, Carotenuto_2022}. Applying the same model to those jets allows us to place independent constraints on the density jump between the interior and exterior of the cavity, but in these cases the value of the internal density still relies on the aforementioned assumption on the external ISM density.
In the following, we discuss how our $\Tilde{E}_0$ solutions, coupled to independent constraints on the jet energy, allow us to place some very valuable and informative constraints on $n_{\rm ISM}$ for jets decelerating in an uniform ISM. Before that, we note that the jets launched from \hh{} during its 2003 outburst are considered to be decelerating in a uniform ISM, and \cite{Steiner_h17} report an effective energy $\Tilde{E}_0 = 1.0^{+2.2}_{-0.7} \times 10^{47} \ {\rm erg}$. Therefore, we choose to include \hh{} in our sample for the following considerations, and the dependence of $E_0$ on $n_{\rm ISM}$ is shown on the fourth panel of Figure \ref{fig:E0_vs_nism}.

While the kinetic energy of these jets cannot be directly measured, it is possible to estimate an independent upper limit on the total energy that the system can provide to the jets during the ejection. In fact, under several assumptions, it is fairly straightforward to estimate an upper limit on the jet energy by considering the power available from accretion during the timescale associated to the ejection, and assuming that no energy is transferred to the jets after the launch.
If the jet is accelerated during a timescale $\Delta t$, we can write for a single component:
\begin{equation}
    E_{\rm jet} \ = \frac{P_{\rm jet} \Delta t}{2} \simeq \frac{1}{2} \eta_{\rm jet}\dot{M} c^2 \Delta t_{\rm obs} = \frac{1}{2} \frac{\eta_{\rm jet} L_X}{\eta_{\rm rad}} \Delta t_{\rm obs},
\label{eq:energy_available_accretion_power}
\end{equation}
where $P_{\rm jet}$ is the jet power in the rest frame of the source, and can be expressed as the fraction $\eta_{\rm jet}$ of the accretion power $\dot{M} c^2$, which can in turn be traced by the simultaneous X-ray bolometric luminosity $L_{X}$ corrected by the radiative efficiency of the accretion flow $\eta_{\rm rad}$ (as $L_{X} = \eta_{\rm rad} \dot{M} c^2$). As generally done in the literature, we assume that the duration of the ejection phase $\Delta t$ should be roughly equivalent to the duration $\Delta t_{\rm obs}$ of the rising phase of the radio flares observed at the moment of launch, which is observable through a radio monitoring with adequate cadence. We note that $\Delta t$ can be shorter than $\Delta t_{\rm obs}$ if the rise is due to synchrotron self-absorption. Furthermore, in Equation \ref{eq:energy_available_accretion_power} we adopt the standard assumption $\eta_{\rm rad} = 0.1$ (e.g.\ \citealt{Frank_King_Raine, Coriat_2012}) and, since we are interested in an upper limit on $E_{\rm jet}$, we consider the (roughly) maximum possible jet power by simply assuming $\eta_{\rm jet} = 1$. We note that $\eta_{\rm jet}$ could also exceed $1$ in presence of a magnetically arrested accretion disk (MAD) and with a contribution from the BH spin (e.g.\ \citealt{Bisnovatyi-Kogan_1974, Narayan_2003, Tchekhovskoy_2011, McKinney_2012, Davis_2020}).

After obtaining an upper limit on the jet energy $E_{\rm max}$, it is possible to combine such information with the constraints on $\Tilde{E}_0$ to place an upper limit on $n_{\rm ISM}$ by:

\begin{equation}
    n_{\rm ISM} < \frac{E_{\rm max}}{\tilde{E}_0 \phi^2} .
    \label{eq:constrain}
\end{equation}

Following this approach, we now discuss the constraints on the external ISM density for each of the sources considered in our sample.

\subsubsection{\maxieight{}}
\label{sec:result_1820}

\setlength{\tabcolsep}{7pt}
\setlength{\extrarowheight}{.7em}
\begin{table*}
\caption{Source and jet parameters used for the calculations discussed in Section \ref{sec:Jet kinetic energy and external ISM density}. From left to right, we list the name of the source, the peak bolometric X-ray luminosity $L_{\rm X}$ used to compute the available accretion power, the derived maximum kinetic energy $E_{\rm max}$ available to the jet, the duration $\Delta t_{\rm obs}$ and the peak flux density $S_{\nu, \rm peak}$ at the frequency $\nu$ of the radio flare used to compute the jet-frame minimum energy $E_{\rm flare}$ from equipartition, and the inferred upper limits on the external $n_{\rm ISM}$ and jet mass $M_0$. Given the multiple sources of uncertainty in the computations of these numbers, we associate a conservative $30\%$ uncertainty to $L_{\rm X}$ and $E_{\rm max}$, and a $50\%$ uncertainty to $E_{\rm flare}$. The references from which these data are obtained are reported in Section \ref{sec:Jet kinetic energy and external ISM density}.}
\label{tab:constraints}
\begin{tabular}{*{11}{l c c c c c c c c c c}}
\hline                                       Source             & $L_{\rm X, flare}$                   &   $E_{\rm max}$       &  $\Delta t_{\rm obs}$  &    $S_{\nu, \rm peak}$  &  $\nu$    & $E_{\rm flare}$                                &   $n_{\rm ISM}$   &   $M_0$\\                             
                   & [erg s$^{-1}$]            			 &  [erg]                 &     [h]              &            [mJy]        &   [GHz]   & [erg]                                          &   [cm$^{-3}$]     &     [g]  \\
\hline                                                                                                                                                                                                                                                        
\maxieight{}       & $2 \times 10^{38}$    &  $ 2  \times 10^{43}$            &  $6.7$ 			    &   		    $50	$           &  $15.5$   &                 $ 9     \times 10^{37}$      &    $ \lesssim 10^{-3}$    & $\lesssim 1.5 \times 10^{22}$  \\
\maxififth{}       & $1 \times 10^{39}$    &  $ 5  \times 10^{44}$            &  $24$    			    &   		$600$           &  $1.3$    &                 $ 1     \times 10^{39}$      &    $ \lesssim 10^{-4}$    & $\lesssim 1 \times 10^{24}$  \\
\xtej{}            & $1 \times 10^{39}$    &  $ 5  \times 10^{44}$            &  $24$    			    &   		$20 $           &  $5.5$    &                 $ 1     \times 10^{37}$      &    $ \lesssim 0.4   $      & $\lesssim 2 \times 10^{23}$  \\
\hh{}              & $3 \times 10^{38}$    &  $ 1  \times 10^{44}$            &  $24$    			    &   	    $35 $           &  $4.8$    &                 $  7    \times 10^{39}$      &    $ \lesssim 10^{-3}$    & $\lesssim 6 \times 10^{22}$\\
\hline                                                                                                                                                              
\end{tabular}                                                                                                                                   
\end{table*}  

Starting with \maxieight{}, we consider the radio flare produced by the ejection and observed on MJD 58306 with AMI-LA, which was characterized by a rising timescale of $6.7$ h and a peak flux density of $\simeq 50$ mJy at $15.5$ GHz \citep{Bright, Homan_qpo}. We also consider the simultaneous X-ray luminosity from \cite{Fabian_2020}, extrapolated using the measured spectral parameters to the $0.5$--$200$ keV energy range using the multi-component feature of the \textsc{webpimms}\footnote{\url{https://heasarc.gsfc.nasa.gov/cgi-bin/Tools/w3pimms/w3pimms.pl}} tool. Using Equation \ref{eq:energy_available_accretion_power}, we obtain $E_{\rm max} \simeq 2 \times 10^{43}$ erg, which is shown as a horizontal dash-dotted orange line in Figure \ref{fig:E0_vs_nism}, and reported in Table \ref{tab:constraints}. Notably, such upper limit is broadly consistent with the jet internal energy $E_{\rm int}$ measured by \cite{Bright}, which obtained values between $10^{41}$ and $10^{43} \ \rm erg$, thanks to a reliable estimation of the size of the jet emitting region 90 days after the jet launch. In addition, \cite{Espinasse_xray} obtained $E_{\rm int} \simeq 5 \times 10^{41}$ erg by resolving the ejecta in the X-ray band with \textit{Chandra} and measuring the broadband radio-to-X-ray spectrum. Both these results were obtained with minimum energy calculations, assuming equipartition between electrons and magnetic fields in the jet plasma \citep{Longair}. 

Using Equation \ref{eq:constrain}, we can place the following constraint on the ISM density surrounding \maxieight{}{}, assuming $\phi = 1\degree$:
\begin{equation}
    n_{\rm ISM, J1820} \lesssim 10^{-3} \ {\rm cm^{-3}},
\end{equation}
with the upper limit being relaxed in case the jet is significantly narrower than $1\degree$ (but unlikely given that the jet appeared to be resolved in one of the X-ray detections reported in \citealt{Espinasse_xray}). We also note that \cite{Tetarenko_2021} measured $\phi = 0.45\degree^{+0.13\degree}_{-0.11\degree}$ for the compact jets in the same source. Interestingly, such result on the $n_{\rm ISM}$ appears to place \maxieight{} in a low-density region filled with hot/coronal-phase ISM \citep{Cox_2005}, similarly to \maxithirt{} and \xte{} \citep{Steiner_xte, Carotenuto_2022}, but in this case the proper motion data can be adequately described with the propagation in a uniform, low-density ISM. If \maxieight{} is in a low-density cavity, it is possible that these jets were not sufficiently energetic to travel up to the cavity \enquote{wall}, if present, or, alternatively, the ISM around \maxieight{} might have a much smoother distribution compared to the two sources mentioned before. 

Finally, we can also obtain a lower limit on the jet kinetic energy, which is represented by the internal energy inferred from the radio flare observed at the moment of ejection, when applying minimum energy calculations (e.g.\ \citealt{Fender2006}). The internal energy $E_{\rm flare, obs}$ in the observer frame is computed considering that the flare spectrum evolves from optically thick to optically thin, under the assumption that such evolution is due to decreasing optical depth to synchrotron self-absorption \citep{Fender_2019_equipartition}. As in most of our cases, if the flare has been monitored at a single frequency $\nu$, we can write:
\begin{equation}
E_{\rm flare, obs} = 1.5 \times 10^{35}  \left(\frac{D}{\rm kpc} \right)^{\frac{40}{17}}\left( \frac{S_{\nu, \mathrm{peak}}}{\rm mJy} \right)^{\frac{20}{17}} \left( \frac{\nu}{\rm GHz} \right)^{-\frac{23}{24}} \ \rm erg,
\label{eq:SSA}
\end{equation}
where $S_{\nu, \mathrm{peak}}$ is the peak flux in mJy and $D$ is the source distance in kpc. We note that this approach has the advantage of being independent of the flare rising timescale, and it can be applied to any flare for which there is evidence for self-absorption and a measurement of the peak flux is available. Most of the flares from BH XRBs detected so far have shown this particular evolution (e.g. \citealt{Tetarenko2017, Russell_1535, Carotenuto2021, Fender_2023}). Adopting the formalism from \citep{Fender_2019_equipartition}, we compute the same energy in the jet frame and we account for the relativistic bulk motion by
\begin{equation}
    E_{\rm flare, RF} = \Gamma_0 E_{\rm flare, obs} \delta^{-\frac{97}{34}},
\label{eq:rest_frame}
\end{equation}
where $\delta$ is the Doppler factor for the approaching jet and we rely on the assumption that $\alpha = 0$ at the time of peak flux \citep{Fender_2019_equipartition}. We further note that this lower limit on $E_0$ gives the lowest jet mass $M_0 = E_0/(\Gamma_0 -1)c^2$ and corresponds to the jet composition of pure $e^\pm$ pairs. For \maxieight{}, using $\Gamma_0, \theta$ from the fit and $S_{nu}, \nu, \Delta t_{\rm obs}$ from the AMI-LA flare, we obtain, through Equation \ref{eq:rest_frame}, $E_{\rm flare, RF} \simeq 9 \times 10^{37}$ erg, which is also shown with a black dotted line in Figure \ref{fig:E0_vs_nism} and reported in Table \ref{tab:constraints}. Given the significant uncertainties associated to this method, we associate a conservative 50\% error to these estimations. We note that, from Figure \ref{fig:E0_vs_nism}, a jet with such kinetic energy would necessarily be propagating in an extremely low-density environment, with $n_{\rm ISM} \ll 10^{-4}$ cm$^{-3}$. It is reasonable to suggest that the $E_{\rm flare}$ and $E_{\rm max}$ lines enclose the physical parameter space for the ejecta, and the true $(E_0, n_{\rm ISM})$ value should lie between the two horizontal lines.

\subsubsection{\maxififth{}}
Similarly to \maxieight{}, we consider the radio flare associated to the jet ejection to estimate the maximum and minimum energies available to the jet. An upper limit on the jet kinetic energy can be obtained with Equation \ref{eq:energy_available_accretion_power}, by considering the radio flare produced by the ejection of the S2 component \citep{Russell_1535}. Such flare, observed with MeerKAT and ATCA on MJD 58017, had an approximate rising timescale of $\simeq 24$ h and a peak flux density of $\simeq 600$ mJy at 1.3 GHz. At the same time, we consider the simultaneous X-ray luminosity from \cite{Tao_2018}, extrapolated using the measured count rates and reported spectral parameters to the $0.5$--$200$ keV energy range using \textsc{webpimms}. We infer a maximum jet energy of $E_{\rm max} \simeq 5 \times 10^{44}$ erg, which, assuming $\phi = 1\degree$, leads to a more stringent constraint on the ISM density surrounding \maxififth{}:
\begin{equation}
    n_{\rm ISM, J1535} \lesssim 10^{-4} \ {\rm cm^{-3}}.
\end{equation}
Such upper limit is roughly one order of magnitude lower than what obtained for  \maxieight{}, and also lower than what inferred for \maxithirt{} and \xte{} \citep{Steiner_xte, Carotenuto_2022, Zdziarski_2023}. This is consistent with the picture of the jets from from \maxififth{} propagating up to a larger distance without the abrupt deceleration observed in the two latter sources.
Considering again the radio flare, we can also use the rising timescale, peak flux density and observing frequency to compute with a minimum energy of $E_{\rm flare, RF} \simeq 10^{39}$ erg in the jet frame, with the same procedure outlined in the previous section and using Equation \ref{eq:rest_frame}. Therefore, the true kinetic energy of the jets from \maxififth{} likely lies in the interval between $10^{40}$ and $10^{44}$ erg, which, for instance, is in line with what estimated for the ejecta from by \grs{} \citep{Mirabel1994, Fender1999, Zdziarski_2014}. All the constraints are shown in Figure \ref{fig:E0_vs_nism} and reported in Table \ref{tab:constraints}.

\subsubsection{\xtej{}}
\label{sec:density_xte}
We apply the same method used above to the jets launched by \xtej{}. First, we consider the radio flare observed with ATCA on MJD 55217, characterized by a rising timescale of $\simeq 24$ h and a peak flux density of $20$ mJy at $5.5$ GHz \citep{Brocksopp}. Including the simultaneous X-ray bolometric luminosity reported in the same paper and extrapolated to the $0.5$--$200$ keV range, we use Equation \ref{eq:energy_available_accretion_power} to compute again the maximum energy available to the jet $E_{\rm max} \simeq 5 \times 10^{44}$ erg, represented in the third panel of Figure \ref{fig:E0_vs_nism}. The application of Equation \ref{eq:constrain} yields the following upper limit on the external ISM density, when assuming $\phi = 1\degree$:
\begin{equation}
    n_{\rm ISM, J1752} \lesssim 0.4 \ {\rm cm^{-3}}.
\end{equation}
which is below the standard galactic $n_{\rm ISM} \simeq 1$ cm$^{-3}$, but does not stringently constrain the environmental density as in the cases of \maxieight{} and \maxififth{}.
On the other hand, given the short distance up to which the jets from \xtej{} propagate (see Figure \ref{fig:angsep_fit_uncertainties_1752}), this result would be consistent with a scenario in which the ejecta have energies similar to the other sources, but travel in a much denser environment. 

\begin{figure*}
\begin{center}
\includegraphics[width=8cm]{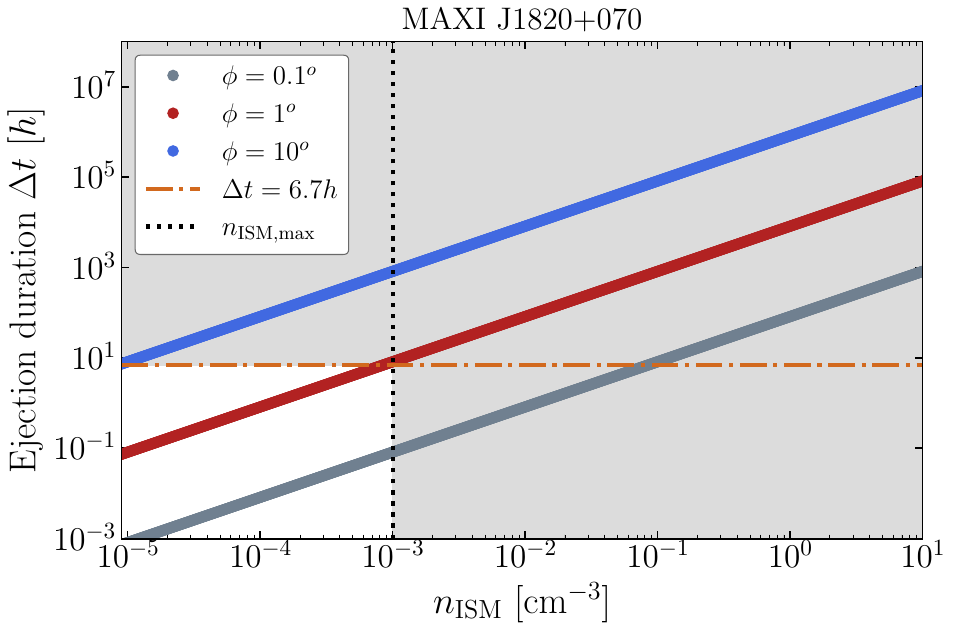}\includegraphics[width=8cm]{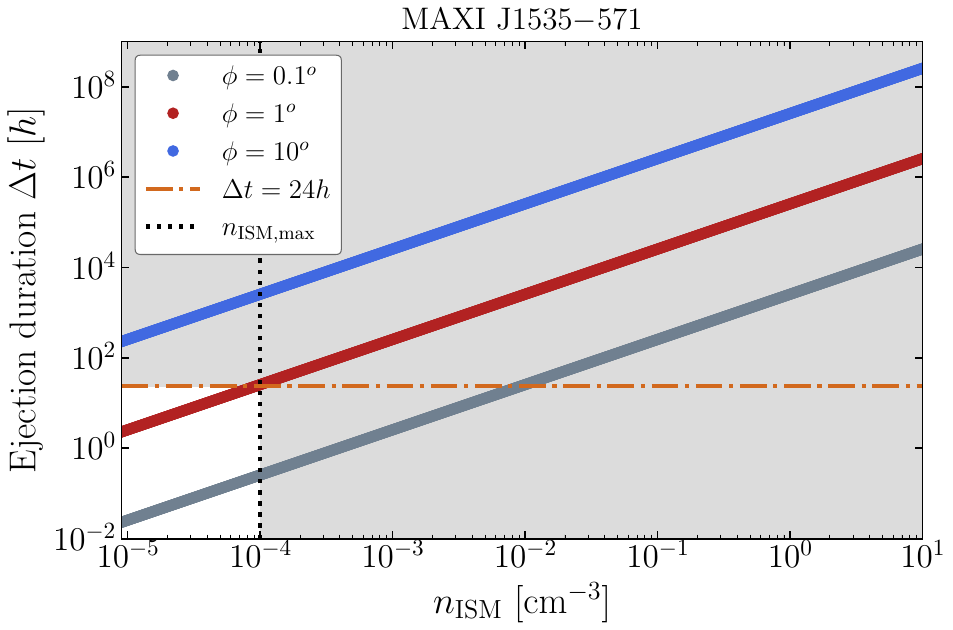}\\ \includegraphics[width=8cm]{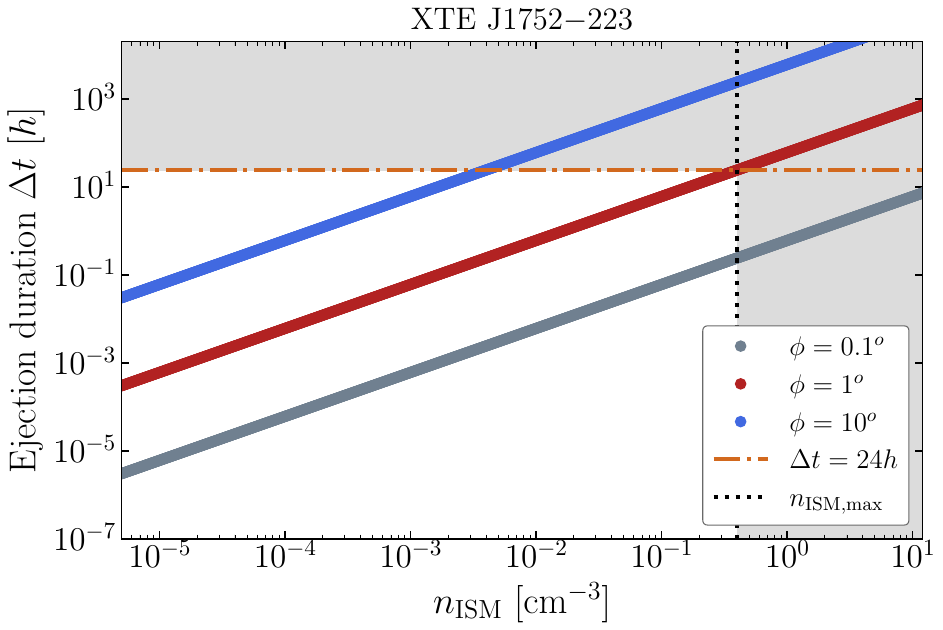}\includegraphics[width=8cm]{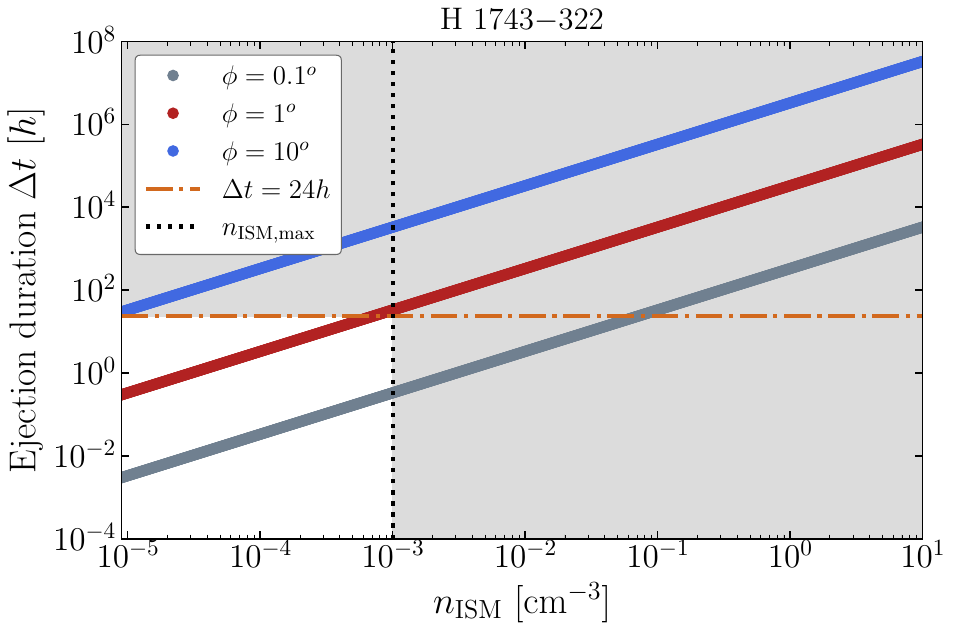}
\caption{Explicit dependence of the ejection duration $\Delta t$ on the external ISM density $n_{\rm ISM}$ for different values of the half-opening angle $\phi$. The dependence is obtained through the constraints on the effective energy $\Tilde{E}_0$, from Equation \ref{eq:energy_available_accretion_power} and from the simultaneous bolometric X-ray luminosity, and it is here shown for the four sources considered in Section \ref{sec:Jet kinetic energy and external ISM density}. For ISM densities above the upper limits presented in Section \ref{sec:Jet kinetic energy and external ISM density}, marked with dotted vertical black lines, the higher energies required would imply the full accretion power to be supplied to the jets over timescales not compatible with the observed flares and state transitions. Regions excluded by our constraints on $\Delta t$ and $n_{\rm ISM}$ are shaded in grey.}
\label{fig:DELTAT_vs_nism}
\end{center}
\end{figure*}

Considering again the radio flare, we combine the rising timescale, peak flux density and observing frequency to estimate a minimum energy of $E_{\rm flare, RF} \simeq 10^{37}$ erg  with the same method as before (and Equation \ref{eq:rest_frame}). Since we do not have a preferred value for the initial Lorentz factor, we assume $\Gamma_0 = 3.5$, consistent with the lower limit from the fit. Again, we associate a conservative $30\%$ uncertainty to this estimation, and we report all values in Table \ref{tab:constraints}. In particular, we mention that the main source of uncertainty is the duration of the ejection $\Delta t$, which could be overestimated in case of a sparse radio monitoring. Nevertheless, the minimum and maximum energies together with the low effective energy appear to point to a denser environment with respect to the other sources. We note that \xtej{} displayed a complex flaring activity during the 2010 outburst, with rapid oscillations between the intermediate and the soft state, and with the likely production of additional, undetected ejecta \citep{Brocksopp}. Hence, it might also be possible that jets from this source are intrinsically less energetic than the single pair of ejecta displayed by \maxithirt{} or \xte{} (with the counter-example of \grs{}, \citealt{Fender_2000}). Such behavior has been also observed in a significant number of other sources, which are inferred to produce multiple subsequent ejecta, which are, however, rarely detected and spatially resolved (e.g.\ \citealt{Homan_2001, Brocksopp_2001, Brocksopp_2002, Fender_2009, Tetarenko2017, Carotenuto2021, Fender_2023}). As of today, it is unclear if these multiple ejecta that should result from the complex flaring activity are not detected due their lower energy content, or for different reasons, possibly linked to the source environment that might be affected by the previous ejecta.

\subsubsection{\hh{}}

Lastly, we consider \hh{} and the radio flare associated to the ejecta that was observed to peak on MJD 52768 with the VLA \citep{McClintock_2009}. The flare had a rising timescale of $\simeq 24$ h, with a peak flux density of $\simeq 35$ mJy at 4.8 GHz. Combining this information with the extrapolated X-ray bolometric luminosity yields a maximum energy available to the jet $E_{\rm max} \simeq 1 \times 10^{44}$ erg (Equation \ref{eq:energy_available_accretion_power}), represented in the fourth panel of Figure \ref{fig:E0_vs_nism}, which then translates through Equation \ref{eq:constrain} to the following upper limit on the ISM density:
\begin{equation}
    n_{\rm ISM, H1743} \lesssim 10^{-3} \ {\rm cm^{-3}}.
\end{equation}
Assuming a half-opening angle of $1\degree$, we confirm the requirement for a low density environment, as already pointed out by \cite{Hao} and \cite{Steiner_h17}.
Considering the radio flare, and again assuming $\Gamma_0 = 3$ and an inclination angle $\theta = 75\degree$  \citep{Steiner_h17}, the jet in its rest frame has a minimum energy of $E_{\rm flare, RF} \simeq 7 \times 10^{39}$ erg from Equation \ref{eq:rest_frame}.

\subsubsection{Ejection duration}

As an alternative way to represent the constraints obtained on the density of the ISM surrounding our sources, we can use Equation \ref{eq:energy_available_accretion_power} to compute the ejection duration associated to the measured $L_{\rm X}$ for any possible value of $E_0$, which is then linked to a specific value of $n_{\rm ISM}$ through the definition of $\Tilde{E}_0$. The results are shown in Figure \ref{fig:DELTAT_vs_nism} for different values of $\phi$. Here, the orange dash-dotted line represents the ejection duration used in the previous subsections to estimate the maximum and minimum energies available to the ejecta. Considering our constraints on $\Tilde{E}_0$, we show that, for \maxieight{}, \maxififth{} and \hh{}, if the jet were to propagate in a denser environment with respect to the upper limits reported in Table \ref{tab:constraints}, the ejection would have required a sustained supply of accretion power $\dot{M}c^2$ over timescales much larger than the ones associated with the radio flares, by orders of magnitude. For instance, if the ejecta from \maxififth{} were propagating in a $10^{-2}$ cm$^{-3}$ ISM with $\phi=1\degree$, to reach the same distance the system should have supplied energy to the jet at the measured rate for more than 1000 h, which is longer than the time interval between the beginning of the outburst and the inferred jet ejection.

It is crucial to remark that these arguments strongly rely on the assumption that the rising timescale of the radio flares is a good proxy for the true ejection duration, to which we have no direct observational access. We note that this is not true for models that consider continuous jets instead of discrete ejections. In a continuous jet model, the resolved radio knots that we observe are believed to be caused by internal shocks between plasma shells accelerated at different speeds \citep{Kaiser, Jamil, Malzac_2013}. While this does not decrease the total amount of energy required for the jet, it relaxes the requirement on the ejection timescale, since the majority of energy is stored in the material of the unseen continuous jet \citep{Kaiser}.

\subsection{BH XRBs in low-density environments}
\label{sec:BH XRBs in low-density environments}

The propagation of the ejecta considered in this work can be adequately described with a single deceleration phase in an homogeneous environment, with an assumed constant ISM density. For three out of the four sources considered in the previous section, combining the upper limits on the maximum available energy with the strong constraints on the effective energy provides us with robust upper limits on the external ISM density, implying that \maxieight{}, \maxififth{} and \hh{} are all harbored in a low-density region in the ISM. For \xtej{}, the results are instead not conclusive. At the same time, the motion and the light curve evolution of the ejecta observed from \maxithirt{} and \xte{} also suggest the presence of low-density ISM cavities in which these sources might be embedded, with internal $n_{\rm ISM} = 10^{-3} \sim 10^{-2}$ cm$^{-3}$ \citep{Hao, Steiner_xte}, and with either a sharp border \citep{Carotenuto_2022}, or a more physical, smooth transition layer \citep{Zdziarski_2023}. Despite this difference, it is remarkable to note that for all of the sources displaying large-scale decelerating jets it has been necessary to invoke an environment with a density up to 4 orders of magnitude lower than the canonical ISM density of 1 cm$^{-3}$. As already argued in \cite{Heinz_2002} for \grs{} and GRO~J1655--40, a low-density environment seems to be a necessary requirement for the jet to propagate up to such large distances (fractions of pc) far from the central compact object. This highlights the importance and the great potential of using the current and future observations of large-scale jets as probes of the environment surrounding BH XRBs.

Despite the emerging scenario, it is currently unclear how such low-density environments might be produced. BH XRBs might be preferentially located in regions occupied by the hot ISM phase (e.g.\ \citealt{Ferriere_2001}), or, more likely, the low-density region/cavity could formed from the feedback of the system itself. The BH surroundings might have been evacuated by the supernova explosion that created the compact object or by a different type of outflow. Several possibilities (with different degrees of plausibility, see discussion in \citealt{Hao}) include the 
winds from the progenitor star (e.g.\ \citealt{Gaensler_2005}), winds from the companion star (e.g.\ \citealt{Sell_2015}) and winds from accretion disk (e.g.\ \citealt{Miller_2006, Fuchs_2006, Munoz_darias_2016}), in addition to the previous activity of the jet itself, whether collimated (e.g.\ \citealt{Gallo_2005, Russell_2007, Heinz_2007, Heinz_2008, Yoon_2011, Coriat_2019}) or, as more recently proposed, uncollimated \citep{Sikora_2023}. Interestingly, such scenario appears to be supported by laboratory experiments testing multiple supersonic plasma ejections in a drift chamber  \citep{Kalashnikov_2021}. After the passage of each ejection, a low-density region can be observed, called \enquote{vacuum trace}, which causes the subsequent ejections to encounter much less environmental resistance in their propagation.

It is worth mentioning that the properties, distribution and chemistry of the ISM around these systems can be investigated through independent observation at different wavelengths, of which a prime example is the mapping of the molecular line emission from the material shocked by the jets, as already done for \grs{}, GRS~1758--258 and 1E~1740.7--2942 \citep{Mirabel_1998, Chaty_2001, Tetarenko_alma, Tetarenko_2020_ism}. Alternatively, it is possible to search for optical H$\alpha$ emission from the same shocked ISM and independently infer its density from the measurement of the integrated luminosity in the diagnostic line and from realistic assumptions on the shock velocity \citep{Dopita_1996, Russell_2007}. Therefore, the importance of these approaches resides also in their potential for partially solving the $E_0$, $\phi$, $n_{\rm ISM}$ degeneracy that is currently present in our model.

\subsection{{Jet mass}}
\label{sec:Jet mass}

\begin{figure*}
\begin{center}
\includegraphics[width=\textwidth]{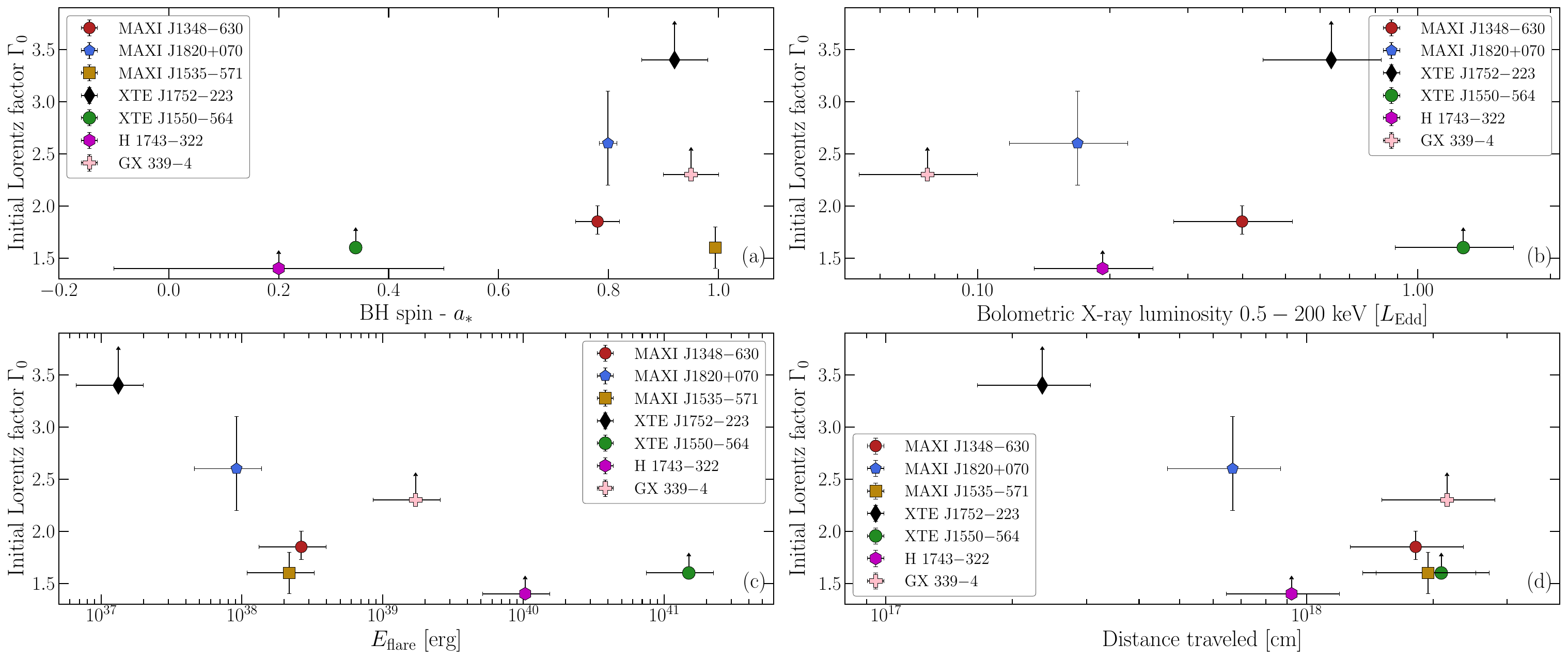}
\caption{Comparison between the inferred initial Lorentz factor $\Gamma_0$ of the ejecta in our sample of large-scale jets and: (\textit{a}) - the dimensionless spin parameter $a_*$, (\textit{b}) - the bolometric X-ray luminosity $L_{\rm X}$ simultaneous to the ejection in Eddington units, (\textit{c}) - the jet frame internal energy $E_{\rm flare}$ inferred from the radio flare associated to each ejection (see text for details), (\textit{d})  the de-projected distance traveled by the jet. We find no clear correlation between $\Gamma_0$ and the parameters shown in the four panels, and more sources are needed to increase the sample of large-scale jets.}
\label{fig:composite}
\end{center}
\end{figure*}

The mass of the ejecta is a parameter of great importance in the study of these systems, and it has not yet been constrained with sufficient accuracy for any source. A mass measurement can directly give us information on the long sought-after composition of the jets from BH XRBs, which is still an open problem. While we have evidence for baryons in the jets from SS~433 from the detection of Doppler-shifted iron emission lines in the X-rays \citep{Kotani_1996, Migliari_2002}, no information is available from the synchrotron spectra of the other jets, both compact and discrete \citep{Fender2006}, although there have been attempts to model compact jets SEDs with hadronic-leptonic models (e.g.\ \citealt{Romero_2005, Pepe_2015, Romero_2017, Kantzas_2021}). In this context, finding evidence for massive ejecta could strongly suggest the presence of cold protons, balanced by a long tail of non-relativistic electrons \citep{Carotenuto_2022}. This is also supported by \cite{Zdziarski_2023}, which argue that a massive jet is unlikely to be pair dominated, and by more recent results reported in \cite{Zdziarski_2024}, which suggest the existence of a fundamental difference in composition between compact jets, pair-dominated, and discrete ejecta, which should instead have a baryonic composition. Evidence of protons in these jets bolster arguments that BH XRBs could represent a class of PeV cosmic ray sources (e.g.\ \citealt{Fender_2005, Cooper_CR}). However, the mechanism for which protons are loaded in the jets remains unclear, as they could be already present at the jet formation or could be entrained during the early phases of the jet motion (see for instance \citealt{ORiordan_2018} and \citealt{Kantzas_2023}).

In this work, we cannot directly provide estimations of the masses of the ejecta for the four sources considered, but we can still place interesting upper limits on it by writing
\begin{equation}
M_{0} < \frac{E_{\rm max}}{(\Gamma_0 - 1)c^2}
\end{equation}
and using the best value of $E_{\rm max}$ and $\Gamma_0$  from the sections above. With the values reported in Table \ref{tab:fit_params_jets}, we find the ejecta launched by \maxieight{} to have a mass
\begin{equation}
M_{0, \rm J1820} \lesssim 1.5 \times 10^{22} \ {\rm g},
\end{equation}
which is equivalent to an upper limit of $\lesssim 8 \times 10^{-12} M_{\odot}$. This estimation is consistent with the $\simeq 10^{20}$ g obtained with minimum energy calculations, assuming one proton per electron \citep{Espinasse_xray}, and it is also consistent with jet masses estimated in other sources with the same method (e.g.\ \citealt{Fender1999, Gallo_2004}). 

Regarding \maxififth{}, we infer a higher upper limit:
\begin{equation}
M_{0, \rm J1535} \lesssim 1 \times 10^{24} \ {\rm g,}
\end{equation}
or $\lesssim 5 \times 10^{-10}  M_{\odot}$. It is likely that the ejecta from \maxififth{} have a larger density contrast than \maxieight{} with the surrounding ISM, as this is probably required in order to propagate a larger distances. Notably, the ejecta from \maxififth{} were unresolved in all the ATCA and MeerKAT detections \citep{Russell_1535}, hence it is not expected to have a volume significantly larger than the jet from \maxieight{}.

Assuming a most likely value $\Gamma_0 = 3.5$, the mass of the jets from \xtej{} is inferred to be
\begin{equation}
M_{0, \rm J1752} \lesssim 2 \times 10^{23} \ {\rm g,}
\end{equation}
equivalent to an upper limit of $\lesssim 1 \times 10^{-10} M_{\odot}$. In this case, while the mass upper limit lies in between what obtained for the two previous sources, the early jet deceleration observed in \xtej{}
likely points towards a denser environment (see Section \ref{sec:Jet kinetic energy and external ISM density}). Similarly to \xtej{}, by assuming $\Gamma_0 = 3$ in the case \hh{}, we can constrain \begin{equation}
M_{0, \rm H1743} \lesssim 6 \times 10^{22} \ {\rm g,}
\end{equation}
equivalent to an upper limit of $\lesssim 3 \times 10^{-11} M_{\odot}$. All the inferred upper limits on $M_0$ are reported in Table \ref{tab:constraints}. 

Assuming a high accretion rate of $\sim$$10^{18}$ g s$^{-1}$ (roughly $0.1 L_{\rm Edd}$), typical of the hard-to-soft state transition \citep{Maccarone}, and assuming that during the ejection the majority of the accreted mass is channeled into the jets, it would take from minutes to hours to accumulate a jet mass in the range $10^{20} \sim 10^{22}$ g. On the other hand, it is generally known that the mass outflow rate for thermal winds can be up to ten times higher than the simultaneous accretion rate (e.g.\ \citealt{Higginbottom_2015, Dubus_2019}). This appears to be consistent with the general picture in which, for BH XRBs, most of the outflow mass is carried by winds, while most of the kinetic feedback is carried by jets \citep{Fender_balance}.

\subsection{The sample of large-scale jets}

After discussing the results of the modeling work presented in this paper, we can consider for the first time the entire sample of large-scale jets detected so far and for which we have information on the source parameters, with the aim of looking for possible
interesting and informative trends/correlations. The current sample includes the three sources considered in this work, namely \maxieight{}, \maxififth{} and \xtej{}, with the addition of \maxithirt{} \citep{Carotenuto2021, Carotenuto_2022}, \xte{} \citep{Sobczak_2000, Hannikainen_2001, Wu_2002, Corbel2002_xte, Steiner_xte} and
\hh{ \citep{McClintock_2009, Steiner_h17}}. We further add \gx{}, that displayed large-scale jets in 2003 and for which, albeit not detecting deceleration, we have a lower limit on the initial Lorentz factor $\Gamma_0 > 2.3$ \citep{Gallo_2004}.

We first compare the initial jet speed ($\Gamma_0$) with the measured dimensionless BH spin $a_*$, as can be seen from panel (\textit{a}) of Figure \ref{fig:composite}. At visual inspection, it is not clear whether there is any evidence of correlation between the two parameters, even if we note that we only have three estimations of $\Gamma_0$ in our sample, while the rest are lower limits. Furthermore, we note that these spin values are obtained with different methods that often yield different results (e.g.\ \citealt{Reynolds_2021, Draghis_2023}). Specifically, the spin of \hh{} is obtained through continuum fitting \citep{Steiner_h17}, while the spins of \maxithirt{}, \xtej{}, \maxififth{} and \gx{} have been obtained with relativistic reflection modeling \citep{Parker_2016, Garcia_2018, Miller_2018, Jia_2022}. Lastly, the spins of \xte{} and \maxieight{} were obtained with the application of the RPM (Relativistic Precession Model, e.g.\ \citealt{Stella_1998, Stella_1999_1, Stella_1999_2}), from \cite{Motta_2014} and \cite{Bhargava_2021}, respectively. From our sample, the BH spin does not seem to have a significant effect on the initial jet speed.

With the aim of comparing the jet speed with the simultaneous accretion rate, we show in panel (\textit{b}) $\Gamma_0$ as a function of the simultaneous bolometric X-ray luminosity $L_{\rm X}$ obtained from the literature and converted to the $0.5-200$ keV energy range with \textsc{webpimms}, as in Section \ref{sec:results_1820}. We plot $L_{\rm X}$ in units of $L_{\rm Edd}$, which is the Eddington luminosity $L_{\rm Edd} \simeq 1.3 \times 10^{39} (M_{\rm BH}/10 \ M_{\odot})$ erg s$^{-1}$ that represents the limit for spherically stable hydrogen accretion \citep{Frank_King_Raine}. To estimate $L_{\rm Edd}$ we need a measurement of $M_{\rm BH}$, but we note that a dynamically confirmed BH mass is only available for \xte{} and \maxieight{} \citep{Orosz_2011, Torres_2020}. In this plot we updated the mass of \maxithirt{}, which was previously estimated from the normalization of the disk blackbody component in the X-ray spectral fitting reported in \cite{Tominaga_1348}, for a non-spinning BH. Adopting the recent spin $a_* = 0.78 \pm 0.02$ measurement by \cite{Jia_2022}, we computed the spin-dependent radius of Innermost Stable Circular Orbit (ISCO) and then use this updated parameter to obtain $M_{\rm BH} = 15.2 \pm 2.3 \ M_{\odot}$ (see \citealt{Tominaga_1348} for the explicit dependence of the mass on the ISCO radius). Moreover, we included the new BH mass estimation $M_{\rm BH} = 12 \pm 2 \ M_{\odot}$ obtained for \hh{} through X-ray reflection spectroscopy (Nathan et al., \textit{submitted}) and the new mass estimation of $M_{\rm BH} = 12 \pm 1 \ M_{\odot}$ obtained for \xtej{} through the analysis of the the soft state and the soft-to-hard spectral state transition \citep{Abdulghani_2024}. As for the previous panel, we do not have enough robust estimations of $\Gamma_0$ and $M_{\rm BH}$ to draw a conclusion, but we note that this plot, after \cite{Fender_belloni_gallo}, is starting be populated and will be highly relevant once more measurements become available.

Another interesting comparison, that we show in panel (\textit{c}) of Figure 
\ref{fig:composite}, can be done between the jet speed and the internal energy inferred from the radio flare observed at the moment of ejection, when applying minimum energy calculations (\citealt{Fender_2019_equipartition}, Equation \ref{eq:SSA}) and converting the internal energy to the jet frame, accounting for the bulk relativistic motion with Equation \ref{eq:rest_frame} (with $50\%$ uncertainties associated). We notice that most flares in our sample had a minimum energy ranging between $10^{37}$ and $10^{40}$ erg. For \hh{} and \xte{}, we assumed $\Gamma_0 = 3$ and inclination angles of, respectively,  $\theta = 75\degree$ \citep{Steiner_h17} and $\theta = 70\degree$ \citep{Steiner_xte}, and for which we obtain $E_{\rm flare, RF} \sim 10^{41}$ erg. We caution that the energies reported in this plot are affected by large uncertainties, both in the peak flux (the peak could be missed in the monitoring or it could be optically thin) of the emitting region and in the conversion from the observer frame to the jet frame. Again, a visual inspection seems to show that there is no clear correlation between these two parameters. In the current understanding, $\Gamma_0$ is the bulk Lorentz factor of the whole jet, while $E_{\rm flare, RF}$ should only result from the relativistic electrons present in the jet plasma. If discrete ejecta have a predominant baryonic composition (e.g. \citealt{Zdziarski_2024}), the mass of the proton will be more important in determining $\Gamma_0$ than the energy contained in the relativistic electrons.

Lastly, we compare the de-projected distance traveled by the jet with $\Gamma_0$, as shown in panel (\textit{d}) of of Figure 
\ref{fig:composite}. Due to the uncertainties on the source distance and jet inclination angle, we assume a conservative 30\% uncertainty on the physical distance traveled by the jet. The initial jet speed does not appear to be a driving factor in determining the distance at which the jet propagates to. We might in fact expect that the distance could be more correlated with the jet mass (or with the jet/ISM density contrast, Savard et al.\ \textit{in prep.}) than with its speed. In this context, a massive jet with a low Lorentz factor will propagate further in a given ISM density than a lighter jet with a higher $\Gamma_0$ and the same kinetic energy.

\section{Conclusions}
\label{sec:Conclusions}

In this paper, we have presented a physical modelling of the motion of the decelerating jets launched by \maxieight{}, \maxififth{} and \xtej{}. Adopting a Bayesian approach, we fitted the jet angular distance data with the dynamical blast-wave model developed by \cite{Wang_model}, and we found that the model provides an excellent description of the jet motion, from the first phase of ballistic motion to the final deceleration phase. In particular, a single Sedov phase in a homogeneous ISM appears to adequately capture the dynamics of the decelerating jets for the entire sample considered. The results obtained from a simple model derived from GRBs demonstrate the high potential of applying some of the well-developed theoretical advancements on GRBs to the jets from BH XRBs. These discrete ejecta can be considered as less-relativistic analogues of GRB jets, with the advantage of providing better access to their physics due to their location in the Galaxy.

From the fits, we are able to place constraints on multiple physical parameters of the jets, including a first estimation of the initial Lorentz factor of the ejecta from \maxieight{} ($\Gamma_0 = 2.6_{-0.4}^{+0.5}$) and estimates for the Lorentz factor ($\Gamma_0 = 1.6_{-0.2}^{+0.2}$), ejection date (MJD $58017.4^{+4.0}_{-3.8}$, soft-intermediate state) and inclination angle $\theta = 30.3\degree^{+6.3\degree}_{-6.3\degree}$ for the ejecta produced by \maxififth{}. By considering the constraints on the effective energies and on the maximum energy available to the jets from the accretion power, we are also able to provide new upper limits on the jet mass and on the ISM density surrounding our sources. Overall, our results support the emerging scenario for which BH XRBs displaying large-scale jets appear to be mostly located in low-density environments. 

Considering the current sample of large-scale jets, we find no clear correlations between the initial jet speed and the BH spin, the simultaneous accretion rate, the jet minimum energy inferred from the flare at the moment of ejection, or the distance traveled by the jet. It is worth mentioning that the lack of evident correlation between most of the parameters considered is nevertheless informative for our current understanding of jets from BH XRBs. While our sample is currently limited to a small number of sources, more observations of decelerating ejecta from BH XRBs are needed in order to test our results, an this will lead us to a significant progress in our understanding of the jet production, propagation, and feedback on the surrounding environment. In this context, two new large-scale decelerating ejecta have been recently discovered in MAXI~J1848--015 \citep{Tremou_2021_ATel, Bahramian_2023} and 4U~1543--47 (Zhang et al.\ \textit{in prep.}), and these jets represents ideal targets for the continuation of this work. In the near future, this approach will be greatly improved by the joint modelling of kinematics and radiation from these ejecta (Cooper et al.\ \textit{in prep.}) and will also benefit from the comparison with the first relativistic hydrodynamic simulations of these objects
(Savard et al.\ \textit{in prep.}).

It is worth to mention that the data currently available for \maxieight{} are of outstanding quality, with a monitoring campaign that covered the first phases of jet motion (VLBI) as well as the final deceleration phase, allowing us to obtain much smaller statistical uncertainties on the model parameters with respect to the other sources. In the coming years, similar monitoring campaigns can be planned and performed in order to cover the entirety of the jet evolution. Milliarcsec-resolution VLBI observations are crucial to observe the jets much closer to the compact object and hence to obtain more stringent constraints on their physical parameters, especially the ejection date \citep{Miller-Jones2019, Wood_2021, Wood_2023}. In the future, the new generation Event Horizon Telescope (ngEHT) will also be able to perform extremely high-resolution (order of $\sim$10 $\mu$as) observations of the jets in the mm range (e.g.\ \citealt{Johnson_2023}). At the same time, dense and long-term monitoring campaigns with sensitive interferometers such as MeerKAT, which already detected a large number of discrete ejecta, but also the future SKA-MID \citep{Braun_2015} and ngVLA \citep{Selina_2018} will be fundamental to follow the deceleration phase and to probe the final phases of the jet evolution.

\section*{Data availability}

The angular separation data for \maxieight{} are published in \cite{Bright, Espinasse_xray} and \cite{Wood_2021}. The same data for \maxififth{} and \xtej{} are available, respectively, in \cite{Russell_1535} and \cite{MJ_2011}. A GitHub repository with  \textsc{jupyter} notebooks and the underlying data to reproduce the analysis in this work is available at \url{https://github.com/f-carotenuto/JetKinematics}.

\section*{Acknowledgments}
We thank the anonymous referee for the feedback which greatly improved the paper.
FC acknowledges support from the Royal Society through the Newton International Fellowship programme (NIF/R1/211296). AJT acknowledges partial support for this work from NASA through the NASA Hubble Fellowship grant \#HST--HF2--51494.001 awarded by the Space Telescope Science Institute, which is operated by the Association of Universities for Research in Astronomy, Inc., for NASA, under contract NAS5–26555. AJT acknowledges the support of the Natural Sciences and Engineering Research Council of Canada (NSERC; funding reference number RGPIN-2024-04458). AAZ acknowledges support from the Polish National Science Center under the grants 2019/35/B/ST9/03944 and 2023/48/Q/ST9/00138, and the Copernicus Academy grant CBMK/01/24. AJC acknowledges support from the Hintze Family Charitable Foundation. This project also made use of \textsc{matplotlib} \citep{matplotlib}, \textsc{numpy} \citep{harris2020array} and Overleaf (\url{https://www.overleaf.com}).




\bibliographystyle{mnras}
\bibliography{modeling_paper} 




\appendix

\onecolumn

\section{Posterior distributions}
\label{sec:Posterior distributions}

We show in this section the corner plots with the posterior distributions of the model parameters for the three sources considered in this work.

\begin{figure*}
\begin{center}
\includegraphics[width=\textwidth]{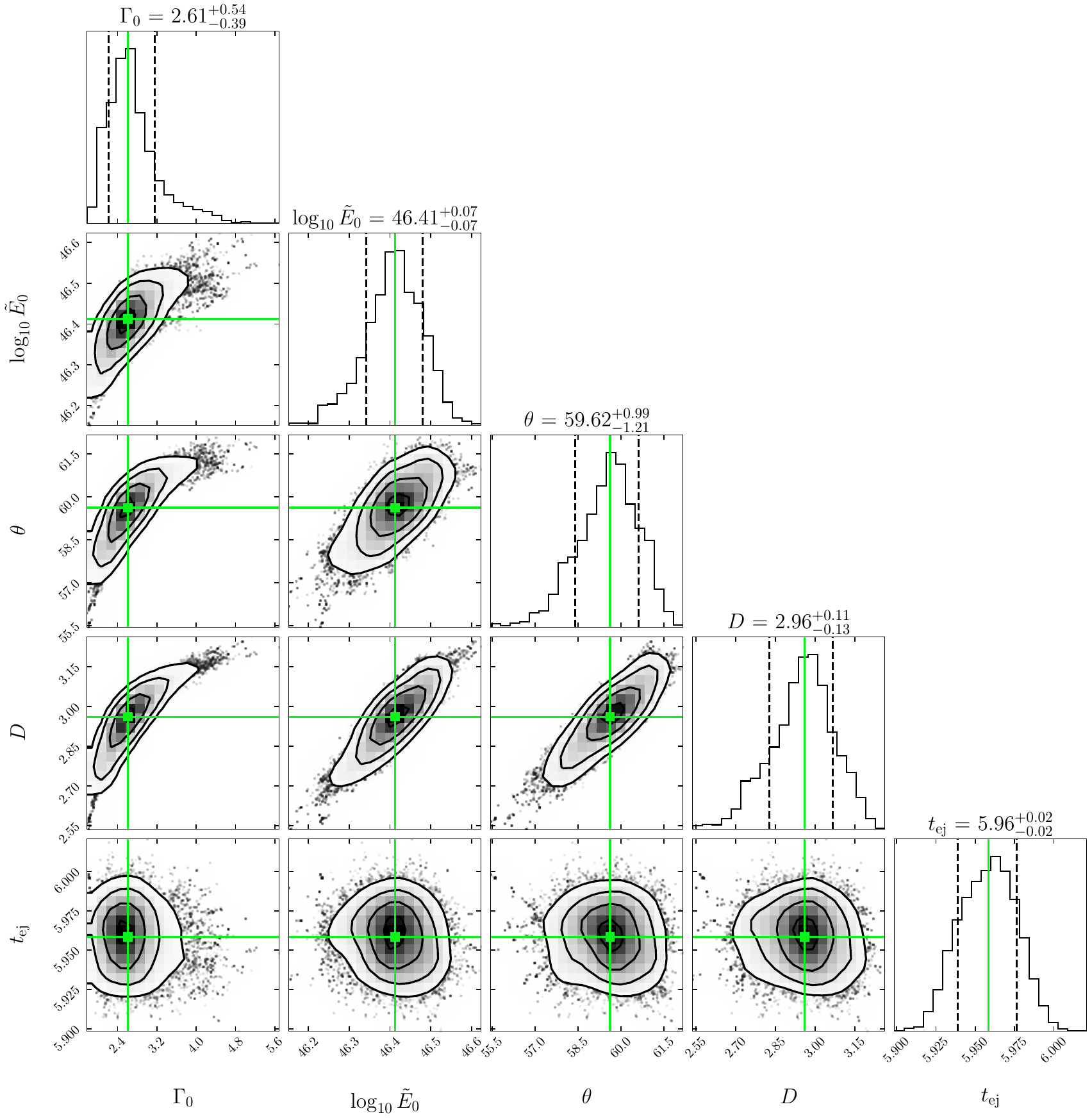}
\caption{Corner plots showing the constraints on the physical parameters of the ejecta from \maxieight{}. The panels on the diagonal show histograms of the one dimensional posterior distributions for the model parameters, including the jet initial Lorentz factor, effective energy, inclination angle and ejection time (here represented as MJD $- 58300$), as well as the source distance. The median value and the equivalent 1$\sigma$ uncertainty are marked with vertical dashed black lines. The other panels show the 2-parameter correlations, with the best-fit values of the model parameters indicated by green lines/squares. The plot was made with the \textsc{corner} plotting package \citep{Foreman_mackey_2016}.}
\label{fig:corner_1820}
\end{center}
\end{figure*}

\begin{figure*}
\begin{center}
\includegraphics[width=\textwidth]{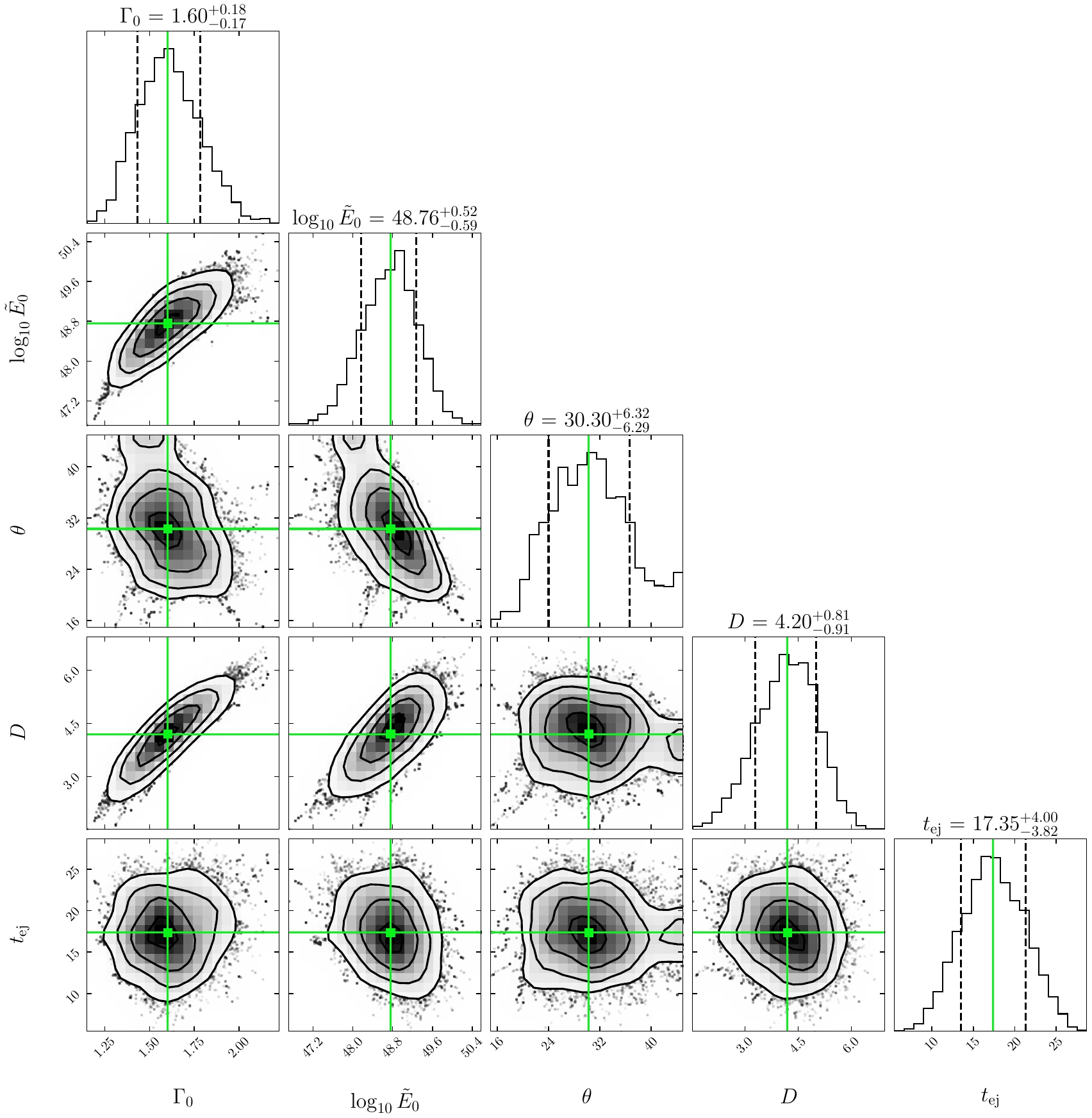}
\caption{Corner plots showing the constraints on the physical parameters of the ejecta from \maxififth{}, same as Figure \ref{fig:corner_1820}. $t_{\rm ej}$ is reported as MJD $- 58000$.}
\label{fig:corner_1535}
\end{center}
\end{figure*}

\begin{figure*}
\begin{center}
\includegraphics[width=\textwidth]{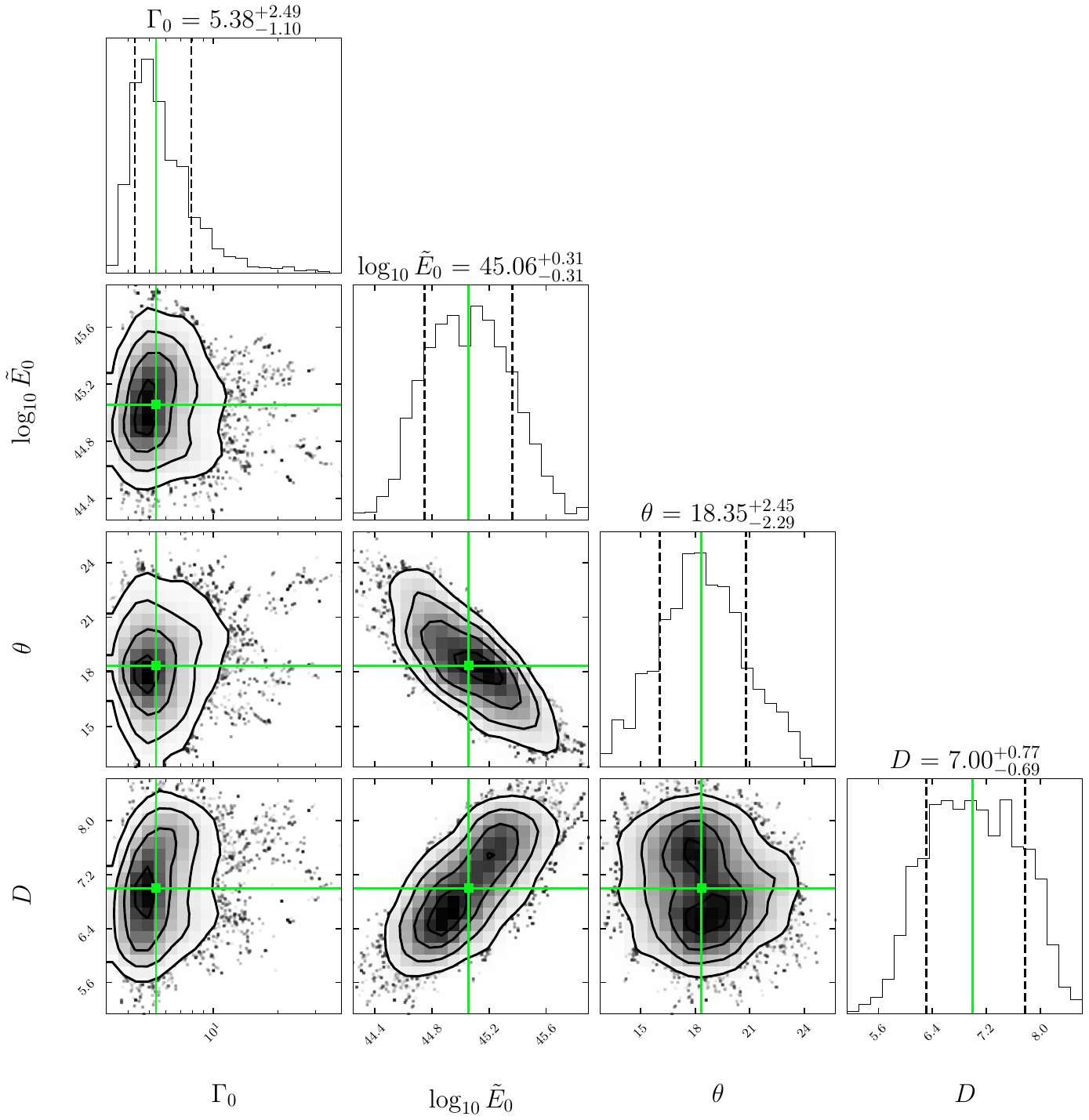}
\caption{Corner plots showing the constraints on the physical parameters of the ejecta from \xtej{}, same as Figure \ref{fig:corner_1820}. Note the log scale for $\Gamma_0$.}
\label{fig:corner_1752}
\end{center}
\end{figure*}

\bsp	
\label{lastpage}
\end{document}